\documentclass[a4paper,prd,noshowpacs,noshowkeys]{revtex4}
\usepackage[latin1]{inputenc}
\usepackage{amsmath}
\usepackage{amssymb}
\usepackage{mathrsfs}
\usepackage{graphicx}
\usepackage{hyperref}

\providecommand{\etaa}[1]{\eta^{\mt{(#1)}}}
\providecommand{\betaa}[1]{\boldsymbol{\eta}^{\mt{(#1)}}}
\providecommand{\rab}[1]{r_{(#1)}}
\providecommand{\rabb}[1]{r_{(\overline{#1})}}

\providecommand{\Labbb}[2]{\Lambda_{(\overline{#1})(\overline{#2})}}
\providecommand{\bsigma}{\boldsymbol{\sigma}}
\providecommand{\bz}{{\mathbf{z}}}
\providecommand{\bx}{{\mathbf{x}}}
\providecommand{\br}{{\mathbf{r}}}

\providecommand{\bkcp}{{\mathbf{k}_{CP}}}
\providecommand{\bn}{{\mathbf{n}}}
\providecommand{\sm}{\mathrm{SM}}
\providecommand{\sulr}[1]{{SU(#1)_L{\otimes}SU(#1)_R}}

\providecommand{\so}[2]{{SO(#1)_{#2}}}

\providecommand{\su}[2]{{SU(#1)_{#2}}}
\providecommand{\gsm}{{SU(2)_L{\otimes}U(1)_Y}}
\providecommand{\uem}{{U(1)_{em}}}
\providecommand{\tphi}{{\tilde{\phi}}}

\providecommand{\oPhi}{{\overline{\Phi}}}
\providecommand{\Iq}{I^{(4)}}
\providecommand{\cS}{{\mathcal{S}}}
\providecommand{\cA}{{\mathcal{A}}}

\providecommand{\ee}{\mathtt{e}}
\providecommand{\pd}{{\otimes}}
\providecommand{\vezes}{{\times}}
\providecommand{\DNN}{{D^{^\mt{\!(2N)}}\!}}
\providecommand{\bM}{{\mathbf{M}}}
\providecommand{\bbM}{{\mathbb{M}}}
\providecommand{\bL}{{\bs{\Lambda}}}

\providecommand{\aver}[1]{\langle #1 \rangle}
\providecommand{\eq}[1]{\begin{equation} #1 \end{equation}}
\providecommand{\eqarr}[1]{\begin{eqnarray} #1 \end{eqnarray}}
\providecommand{\mss}[1]{\mbox{\scriptsize $#1$}}
\providecommand{\mt}[1]{{\mbox{\tiny $#1$}}}
\providecommand{\hs}[1]{\hspace{#1}}
\providecommand{\bs}[1]{\boldsymbol{#1}}
\providecommand{\ponto}{{\cdot}}
\def\Tr{\mathrm{Tr}}
\providecommand{\tp}{{\mss{\mathsf{T}}}}
\providecommand{\diag}{\mathrm{diag}}
\providecommand{\re}{\mathrm{Re}}
\providecommand{\im}{\mathrm{Im}}
\providecommand{\RR}{\mathbb{R}}
\providecommand{\CC}{\mathbb{C}}

\providecommand{\dcol}[2]{\par\smallskip
\begin{minipage}[T]{.4\linewidth}
#1
\end{minipage}
\begin{minipage}[T]{.55\linewidth}
#2
\end{minipage}
\smallskip
}

\font\bb=bbmss10 scaled 970
\def\id{\mbox{\bb 1}}
\begin{document}
\title{
Custodial $SO(4)$ symmetry and CP violation
in N-Higgs-doublet potentials
}
\author{C.~C.~Nishi}
\email{celso.nishi@ufabc.edu.br}
\affiliation{
Universidade Federal do ABC - UFABC\\
Rua Santa Adélia, 166, 09.210-170, Santo André, SP, Brazil
}

\begin{abstract}
We study the implementation of global $SO(4)\sim SU(2)_L\otimes SU(2)_R$ symmetry in
general potentials with N-Higgs-doublets in order to obtain models with custodial
$SO(3)_C$ symmetry. We conclude that any implementation of the custodial $SO(4)$
symmetry is equivalent, by a basis transformation, to a canonical one if $SU(2)_L$
is the gauge factor, $U(1)_Y$ is embedded in $SU(2)_R$ and we require $N$ copies of
the doublet representation of $SU(2)_R$.
The invariance by $SO(4)$ automatically leads to a CP-invariant
potential and the basis of the canonical implementation of $SO(4)$ is aligned to a
basis where CP symmetry acts in the standard fashion.
We show different but equivalent implementations for the 2-Higgs-doublets model,
including an implementation not previously considered.

\end{abstract}
\maketitle
\section{Introduction}
\label{sec:intro}

Despite the enormous success of the Standard Model of Particle Physics (SM) in
accounting for all the known experimental data on high energy physics to
date, several theoretical and observational ingredients pressure the SM
for extensions. One can list naturalness problems such as the hierarchy problem or
observational facts such as the need for new sources of CP violation, the need for
nonordinary weakly interacting stable matter (dark matter) or a natural explanation
for the smallness of nonzero neutrino masses.

The simplicity of the scalar sector of the SM, however, has some unique features
owing to its minimality. Because of the presence of one Higgs doublet, the Higgs
potential possesses a global accidental $\so4{}$ symmetry larger than the usual
electroweak (EW) gauge symmetry $\su2{}\pd U(1)_Y$\,\cite{veltman.97}. 
Even after electroweak symmetry breaking (EWSB),
through the only symmetry breaking pattern, the $\so3C$ portion of the global
symmetry, usually called custodial
symmetry\,\cite{georgi:EFT,sikivie:NPB80,willenbrock}, remains in the potential. The
symmetry is only approximate because it is respected by the gauge interactions only
in the $g_Y\to 0$ limit and by the Yukawa interactions only if the quarks in the same
doublet have degenerate masses (it is then related to isospin symmetry).
Nevertheless, such approximate symmetry guarantees the gauge bosons $W_a^\mu$
transform as a triplet of $\so3C$ in the $g_Y\to 0$ limit and explains the absence of
large radiative corrections to the $\rho$ parameter from its tree-level value
$\rho^{(0)}=M^2_W/M^2_Z\cos^2\theta_w=1$, hence the name custodial symmetry.

When we try to extend the SM, the most simple extensions of the SM consider some
enlargement of the scalar sector of the theory since it is the least verified sector
either because the only elementary scalar of SM, the physical Higgs boson, has
escaped discovery so far or we have had no final confirmation about the mechanism
behind EWSB\,\cite{isidori:2009} 
and the origin of all the masses of the elementary particles of the SM.
The simplest way to extend the scalar sector is to consider the replication of the
Higgs doublet, leading to the N-Higgs-doublet models (NHDMs).
The minimal version called 2-Higgs-doublets model (2HDM), was originally considered
to implement the spontaneous CP violation mechanism\,\cite{lee:scpv}, and more
recently as simple extensions to address problems such as the little hierarchy
problem\,\cite{barbieri.hall}, the presence and stability of dark matter
candidates\,\cite{barbieri.hall,IDM:dark}, and to provide new sources of CP
violation when extended to a 3HDM\,\cite{osland}.
This model has also been extensively studied in the last years as the effective
scalar sector of the MSSM\,\cite{pilaftsis,carena,haber.silva:susy}.
For example, Ref.\,\onlinecite{pilaftsis} presents a detailed study of the effective
2HDM potential that arises in the MSSM with explicit CP violation, after including
radiative corrections.

Another reason to study this type of model is the presence of a horizontal space
formed by the $N$-Higgs-doublets that possess the same gauge quantum numbers. An
$\su{N}H$ transformation in such space, then, only amounts to a reparametrization of
the potential\,\cite{botella.silva,GH,haber:I.II,nhdm:cp,maniatis:1}. At the same
time that the presence of a horizontal space makes the model more complex and less
predictable due to the profusion of allowed couplings it also allows the study of
the potential and its possible symmetry breaking patterns through the study of the
orbit space of the gauge
invariants\,\cite{nhdm:V,ivanov:mink,nhdm:orbit,nhdm:orbit2}.

In the NHDM extensions, with $N>1$, the global $\so4{}$ symmetry does not arise
naturally as an accidental symmetry but has to be imposed. Once it is imposed and
leads to the custodial symmetry $\so3C$, it can play its original role to protect
large radiative corrections to the $\rho$
parameter\,\cite{pomarol:rho,herquet:prl,maniatis:so4,haber:III}.
In the context of the 2HDM, the global $\so4{}$ symmetry leads to the degeneracy of
the charged scalar and one of the neutral scalars\,\cite{scalar.rho:cancel}.
In one version\,\cite{herquet:prl}, it is possible to construct a 2HDM model with
custodial symmetry such that the state degenerate to the charged Higgs is a CP-even
neutral Higgs, $M_{H^0}=M_{H^\pm}$, instead of the more usual $M_{A^0}=M_{H^\pm}$
scenario.

We intend to consider here the NHDM extensions of the SM possessing the custodial
$\so4{}$ symmetry which, in turn, can be broken down to the custodial $\so3C$
symmetry that protects the $\rho$ parameter.
Larger global symmetries can be considered\,\cite{deshpande.ma} but in general it
will produce more pseudo-Goldstone bosons\,\cite{pseudo-G} in addition to the usual
three would-be Goldstone bosons absorbed by the Higgs mechanism.
We can separate our goals in this study into two categories: construction and
identification.
The first refers to the construction of general custodial invariant NHDM potentials
and analysis of the models.
The second goal refers to the methods of identifying the $\so4{}$ symmetry if it is
not manifest and concealed such as the CP symmetry acting in a general
basis\,\cite{botella.silva,GH,nhdm:cp}.
It is then important to study the interplay between the imposition of $\so4{}$
symmetry and the reparametrization group $\su{N}H$. In particular, it is
important to emphasize that the global $\so4{}$ symmetry, like CP symmetry, is not
contained in $\gsm\pd\su{N}H$ as most of the symmetries usually considered for NHDMs
are.
See, for instance, the symmetries for 2HDM and 3HDM in
Refs.\,\cite{ivanov:mink,silva:symmetries,nhdm:orbit2}.
The analysis of basis transformations is also crucial when extending some symmetries
of the potential to the Yukawa interactions, since some symmetries of the potential
might be broken by the Yukawa terms as well as the
converse\,\cite{pomarol:rho,haber:III,silva:yukawa1,maniatis:yukawa}.

In the context of the 2HDM, many studies\,\cite{pomarol:rho,haber:III,maniatis:so4}
show that CP invariance and custodial symmetry are intimately connected. In fact, we
will see that such connection can be extended to general custodial-invariant NHDMs.
In certain cases the knowledge of the presence of the CP symmetry and how
it acts on the fields, allows us to infer if $\so4{}$ is also present as a
symmetry.
We can also borrow the various methods of identifying CP invariance independently of
the basis\,\cite{botella.silva,haber:I.II,nhdm:cp}.
Compared to the SM, where global $\so4{}$ symmetry, custodial $\so3C$
symmetry and CP symmetry are automatic in the scalar sector,
we will see that a NHDM potential invariant by $\so4{}$ symmetry and a
potential invariant by CP symmetry differ only minimally in an easily identifiable
manner for $N=2$ and $N=3$ Higgs doublets.
For $N\ge 4$, a custodial-invariant potential is also more constrained than 
a CP-invariant potential but the difference is more involved.

The paper is organized as follows: 
In Sec.\,\ref{sec:canonical} we study one particular implementation of the global
custodial $\so4{}$ symmetry called canonical implementation and how it is related to
the CP symmetry and basis transformations.
In particular, in Sec.\,\ref{sec:equiv} we show any implementation of the global
symmetry $\so4{}$ is equivalent to the canonical implementation.
Section \ref{sec:2hdm} shows the various implementations for the 2HDM considered in
the literature, shows their equivalence to the canonical implementation and also
shows one different but equivalent implementation.
We discuss some consequences of this work in Sec.\,\ref{sec:consequences}, followed
by the conclusions.
To clarify the nomenclature regarding the various groups involved in the study of 
the custodial symmetry, e.g., $\so3{}$, $\so4{}$ and $\sulr{2}$, we refer to
Appendix \ref{subsec:notation}, where we list the group/subgroup
structure of the global symmetries involved in the SM.
Instead of distinguishing the various (custodial) symmetries by names, we will denote
them by their group structure and use the isomorphic groups interchangeably, often
considering their local structure only. We will often include a subscript ``$C$''
(custodial) to denote the groups, as in $\so4C$.

\section{Canonical implementation of $\so4C$ in NHDM}
\label{sec:canonical}

Let us consider $N$ Higgs-doublets of $SU(2)_L$ of hypercharge $Y=1$:
$\phi_a$, $a=1,\ldots,N$.

Let us also define a $2\vezes 2$ complex matrix of
fields\,\cite{willenbrock,isidori:2009}
\eq{
\label{Phia}
\Phi_a\equiv(\tilde{\phi}_a|\phi_a)
=\begin{pmatrix}
\phi_a^{0*} & \phi_a^+ \cr 
-\phi_a^- & \phi_a^0
\end{pmatrix}
\,,
}
where $\tilde{\phi}_a\equiv\epsilon\phi_a^*$ ($\epsilon\equiv i\sigma_2$) defines the
tilde ($\tilde{~}$) operation on a doublet which transforms a $Y=1$ doublet to a
$Y=-1$ doublet. We can also define the $(\tilde{~})$ operation on any $2\times 2$
complex matrix $h$ as
\eq{
\label{def:tilde:22}
\tilde{h}\equiv \epsilon h^*\epsilon^\dag\,.
}
It is straightforward to see that such operation is trivial on $\Phi_a$ of
Eq.\,\eqref{Phia}, i.e.,
\eq{
\label{tPhi=Phi}
\tilde{\Phi}_a=\Phi_a\,.
}

We can define the Canonical Implementation (CI) of the custodial group $\sulr{2}$ by
the universal action upon $\Phi_a$, $a=1,\ldots,N$, 
\eq{
\label{Phi:LR}
\Phi_a\to U_L\Phi_a U_R^\dag\,,\qquad 
U_L\in \su2L \text{ and } U_R\in \su2R.
}
The action \eqref{Phi:LR} sets the representation of $\Phi_a$ as $(2,2)$ under the
custodial group $\sulr2$. Therefore, each Higgs-doublet $\phi_a$ has the same
transformation property under $\sulr{2}$ as the SM Higgs-doublet.

Within this section we will focus on the canonical implementation of the
custodial group $\so4C$ and any mention to the custodial group will denote this
implementation. Within the 2-Higgs-doublet model, this implementation was named type
I in an early work\,\cite{pomarol:rho}.

It is clear that the identity \eqref{tPhi=Phi} preserves the action \eqref{Phi:LR}
since 
\eq{\nonumber
\epsilon U^*\epsilon^\dag=U\,,
\quad\text{for any $U\in \su2{}$}.
}
Using the operation \eqref{def:tilde:22} we can write the previous identity as
\eq{
\label{tU=U}
\tilde{U}=U\,,
\quad\text{for any $U\in \su2{}$}.
}

If the transformation \eqref{Phi:LR} is promoted to a symmetry of the
$N$-Higgs-doublet potential, we say the potential is \textit{custodial-invariant}.
As in the SM, such symmetry is explicitly broken by the gauge interactions that are
invariant only by the gauged $\gsm$ subgroup. 
In the usual EWSB scenario, the three would-be Goldstone bosons that appear in the
breaking of $\sulr{2}\to\su2{L+R}$ coincide with the usual would-be Goldstone bosons
of the breaking of $\gsm\to \uem$. They are then absorbed as the longitudinal
components of the gauge bosons $W^{\pm},Z^0$ by the Higgs mechanism, such that no
physical Goldstone bosons appear.
Other symmetry-breaking patterns exist where physical pseudo-Goldstone bosons
appear. That is the case of certain 2-Higgs-doublet potentials with global
symmetries equal or larger than $\so4C$ (see Sec.\,\ref{subsec:vevs} and
Ref.\,\onlinecite{deshpande.ma}).

The correspondence between the representation $(2,2)$ of $\sulr{2}$ and the
vector representation $\bs{4}$ of $\so4C$ guarantees we can construct a
4-vector out of the real and imaginary parts of the two components of $\phi_a$.

The usual correspondence is found by defining four real fields
$\etaa{a}=(\etaa{a}_\mu)=(\etaa{a}_0,\betaa{a})$,
$\betaa{a}=(\etaa{a}_1,\etaa{a}_2,\etaa{a}_3)$, from the relation
\eq{
\label{Phi->eta}
\Phi_a = \etaa{a}\id +i\bsigma\ponto\betaa{a}
=
\begin{pmatrix}
\etaa{a}_0 + i\etaa{a}_3 & i\etaa{a}_1 + \etaa{a}_2 \cr
i\etaa{a}_1 - \etaa{a}_2 & \etaa{a}_0 - i\etaa{a}_3
\end{pmatrix}
\,.
}
We can rewrite \eqref{Phi->eta} in a more compact form
\eq{
\label{Phi->eta:2}
\Phi_a = \etaa{a}_\mu \ee_\mu\,,
}
by defining 
$
(\ee_\mu)\equiv(\id,i\sigma_1,i\sigma_2,i\sigma_3)\,.
$

The custodial symmetry $SU(2)_L\pd SU(2)_R$ acting as \eqref{Phi:LR} induces
in $\etaa{a}$ the transformations
\eq{
\label{eta:SO4}
\etaa{a}_\mu\to R_{\mu\nu}(U_L,U_R)\etaa{a}_\nu\,,
\qquad R(U_L,U_R)\in SO(4)\,,
}
where the convention of summation of repeated indices is implicit for
$\mu,\nu=0,1,2,3$.
It can be shown that $R(U_L,U_R)$ defines a two-to-one mapping between
$\sulr2$ and $\so4C$ with the explicit function
\eq{
R_{\mu\nu}(U_L,U_R)\equiv\tfrac{1}{2}
\Tr[\ee_\mu^\dag U_L\ee_\nu U_R^\dag]\,.
}
Hence $\etaa{a}$ can be regarded as vectors in a four-dimensional euclidean space. 
We can also recover the $\su2{L+R}\to\so3{}$ mapping for $U_R=U_L$\,\cite{nhdm:cp}.

We can also define a 4-component complex vector (in a four-dimensional real vector
space) that transforms more directly under $\sulr{2}$.
Let us define
\eq{
\label{def:chi}
\chi_a\equiv
\begin{pmatrix}
\phi_a \cr -\tphi_a
\end{pmatrix}
\,.
}
The action \eqref{Phi:LR} is then translated into
\eq{
\label{chi:LR}
\chi_a\to 
U_R{\otimes}U_L\, \chi_a\,.
}
The transformation property \eqref{chi:LR} can be deduced if we rewrite
\eqref{Phi:LR} as $\Phi_{aij}\to U_{Lii'}U^*_{Rjj'}\Phi_{ai'j'}$ [note
that\linebreak
$\chi_a=(\epsilon{\otimes}\id_2)(\Phi_{a11},\Phi_{a21},\Phi_{a12},\Phi_{a22})^\tp$
implies the transformation property $\chi_a\to(\epsilon{\otimes}\id_2)
(U_R^*{\otimes}U_L)(\epsilon^\dag{\otimes}\id_2)\chi_a$] and finally use  
the identity \eqref{tU=U}.

One can also write explicitly the linear relation between $\chi_a$ and
$\etaa{a}$ as
\eq{
\label{chi->eta}
\chi_a=\mathcal{B}\etaa{a}\,,
}
where the transformation matrix $\mathcal{B}$ can be read off from \eqref{Phi->eta}.
With this change of basis it is possible to write explicitly how the generators of
$\sulr{2}$ acting on $\chi_a$ are related to the generators of $\so4C$ acting on
$\etaa{a}$. Such generators can be found in appendix \ref{ap:chi->eta}.

\subsection{Custodial invariants and explicit CP invariance}
\label{subsec:Cinv}

For the construction of a renormalizable N-Higgs-doublet potential that is invariant
by the custodial group $\so4C$ in the canonical implementation \eqref{Phi:LR},
we need to identify the minimal custodial invariants.

The terms that are invariant by \eqref{Phi:LR}, up to mass order 4, are
\eqarr{
\label{su2lr:inv}
\Tr[\Phi_a^\dag\Phi_b]&=&\phi_a^\dag\phi_b+h.c.\,,\cr
\Tr[(\Phi_a^\dag\Phi_b)(\Phi_c^\dag\Phi_d)]&=&
\phi_a^\dag\phi_b\phi_c^\dag\phi_d
-\phi_a^\dag\phi_c\phi_b^\dag\phi_d
+\phi_a^\dag\phi_d\phi_b^\dag\phi_c
\cr &&
+\ h.c.\,
}
It was necessary to use some reordering identities such as
$\phi_a^\dag\tilde{\phi}_b\tilde{\phi}_c^\dag\phi_d=
\phi_a^\dag\phi_d\phi_b^\dag\phi_c-\phi_a^\dag\phi_c\phi_b^\dag\phi_d$
by making use of the identity
$
\epsilon_{ij}\epsilon_{kl}=\delta_{ik}\delta_{jl}-\delta_{il}\delta_{jk}\,.
$
There are no other invariants of the same orders and apparently independent
invariants are reducible to the ones in Eq.\,\eqref{su2lr:inv}. For instance,
$\det\Phi_a=\tfrac{1}{2}\Tr[\Phi_a^\dag\Phi_a]$, and
$\epsilon_{ii'}\epsilon_{jj'}\Phi_{aij}\Phi_{bi'j'}
=\tfrac{1}{2}\Tr[\Phi_a^\dag\Phi_b]$, 
due to \eqref{tPhi=Phi}.

We immediately notice that the custodial invariants \eqref{su2lr:inv} are
invariant by the canonical CP (CCP) transformations
\eq{
\label{def:CCP}
\phi_a(x_0,\bx)\stackrel{\rm CCP}{\to} \phi_a^*(x_0,-\bx), \quad a=1,\ldots,N.
}
Such transformations act linearly upon
$\chi_a$\,\eqref{def:chi} as
\eq{
\label{chi:CCP}
\chi_a\stackrel{CCP}{\to} 
\begin{pmatrix}
0 & \epsilon \cr 
-\epsilon & 0 
\end{pmatrix}
\chi_a\,,
}
and upon $\etaa{a}$\,\eqref{Phi->eta} as
\eq{
\label{eta:CCP}
(\etaa{a}_0,\etaa{a}_1,\etaa{a}_2,\etaa{a}_3)
\stackrel{CCP}{\to} 
(\etaa{a}_0,-\etaa{a}_1,\etaa{a}_2,-\etaa{a}_3)\,.
}
We realize that $\etaa{a}_0,\etaa{a}_2$ are CP-even fields while
$\etaa{a}_1,\etaa{a}_3$ are CP-odd fields.
Such transformation is in accordance with
$\Phi_a\to\Phi_a^*=\epsilon\Phi_a\epsilon^\dag$.

Since any custodial-invariant potential can be written as functions of the
basic custodial invariants \eqref{su2lr:inv}, we can conclude that 
\textit{\!
a $N$-Higgs-doublet potential which is invariant by the custodial group $\so4C$ in
its canonical form \eqref{Phi:LR} is automatically explicitly CP-invariant in its
canonical form \eqref{def:CCP}.\!
}
This result was established in Ref.\,\onlinecite{pomarol:rho} for the 2HDM potential
considering the canonical implementation (type I) as well as a different
implementation named type II. The latter, in fact, can be shown to be equivalent to
the canonical implementation in a different basis\,\cite{maniatis:so4,haber:III}.
We will discuss basis transformations in Sec.\,\ref{sec:basischange}
and show in Sec.\,\ref{sec:equiv} that any implementation of $\so4C$ in a
NHDM potential is equivalent to the canonical implementation in some basis.

We can ask the converse: Is an explicitly CP invariant NHDM potential
automatically invariant by the custodial group $\so4C$? We know from the 2HDM
potential\,\cite{pomarol:rho,maniatis:so4,haber:III} that the answer is
negative. The immediate following question is: what additional feature
makes a CP-invariant potential also custodial-invariant?
We will show in the following, within the canonical implementation, that the
additional condition is minimal for $N\le 3$.
The characterization for $N\ge 4$ will be seen to be more involved.

For the 2HDM potential, the relation between custodial invariance and CP symmetry
was already explored in Refs.\,\cite{maniatis:so4,haber:III}.
When basis transformations are allowed, such a relation can be related to the
existence of a basis where the implementation of both the
custodial group $\so4C$ and the CP symmetry are the canonical ones.
We will generalize this notion to general N-Higgs-doublets.

To distinguish between a NHDM potential invariant by the
CCP\,\eqref{def:CCP} and by the custodial group $\so4C$\,\eqref{Phi:LR} we need to
analyze the terms allowed by each symmetry to appear in a general potential.

We know the most general renormalizable NHDM potential can be written in terms of
the linear and quadratic combinations of the minimal gauge invariants\,\cite{nhdm:cp}
\eq{
\label{phi:ab}
\phi^\dag_a\phi_b\,,\quad
a,b=1,\ldots,N.
}
Only $4N-4$ of them are algebraically independent\,\cite{nhdm:orbit} when the
fields are considered as c-numbers. This fact affects the minimization procedure.

Instead of working with complex quantities, we can consider the equivalent
set of real quantities
\eqarr{
\label{re:im}
\re(\phi_a^\dag\phi_b),&~~& a\le b=1,\ldots,N, \cr
\im(\phi_a^\dag\phi_b),&~~& a< b=2,\ldots,N.
}

If we succeed in writing the gauge invariants \eqref{re:im} in terms of the 
custodial invariants \eqref{su2lr:inv}, then we can write a general
custodial invariant potential.

For the terms of mass order 2, we can check the correspondence is
\eqarr{
\label{phi->Phi:re}
\re(\phi_a^\dag \phi_b)&=&\tfrac{1}{2}\Tr[\Phi_a^\dag\Phi_b]
\,,\quad a\le b=1,\ldots,N, \\
\label{phi->Phi:im}
\im(\phi_a^\dag \phi_b)&=&\tfrac{i}{2}\Tr[\Phi_a^\dag\Phi_b\sigma_3]
\,,\quad a<b=2,\ldots,N.
}

The term \eqref{phi->Phi:im}, however, is not custodial-invariant but transforms as
the third component of a vector $\bz_{ab}=(z^{A}_{ab})$ of $\su2R$ defined
as
\eq{
\label{zi:def}
z^A_{ab}\equiv \tfrac{i}{2}\Tr[\Phi_a^\dag\Phi_b\sigma_A]\,,\quad
A=1,2,3\,.
}
In fact, the components 1 and 2 of $\bz_{ab}=(z_{ab}^A)$ are not $U(1)_Y$ invariants
as
\eq{
z^1_{ab}=\im(\phi_a^\tp \epsilon \phi_b)\,,\quad
z^2_{ab}=\re(\phi_a^\tp \epsilon \phi_b)\,.
}
Notice $\bz_{aa}=\bs{0}$ and there is no such term in the SM ($N=1$).

Both custodial invariance and CCP symmetry forbid any terms of the form below
in the potential:
\eqarr{
\label{absent:im}
\im(\phi_a^\dag \phi_b)\,,&~&
a<b=2,\ldots,N,
\\
\label{absent:re.im}
\re(\phi_a^\dag \phi_b)\im(\phi_c^\dag \phi_d)\,,&~&
a\le b=1,\ldots,N,~c<d=2,\ldots,N,\qquad
}

On the other hand, both custodial invariance and CCP symmetry allow the terms
\eqarr{
\label{present:re}
\re(\phi_a^\dag \phi_b)\,,&~&
a\le b=1,\ldots,N,
\\
\label{present:re.re}
\re(\phi_a^\dag \phi_b)\re(\phi_c^\dag \phi_d)\,,&~&
a\le b=1,\ldots,N,~c\le d=1,\ldots,N.\qquad
}

The only missing fourth-order terms are CCP-invariant but not custodial-invariant.
They are
\eq{
\label{absent:im.im}
\im(\phi_a^\dag \phi_b)\im(\phi_c^\dag \phi_d)\,,\quad
a<b=2,\ldots,N,~c<d=2,\ldots,N\,.
}
However, if combined with other terms as
\eq{
\label{z.z}
\bz_{ab}\ponto\bz_{cd}=
\im(\phi_a^\dag\phi_b)\im(\phi_c^\dag\phi_d)
+\re(-\phi_a^\dag\phi_d\phi_b^\dag\phi_c +\phi_a^\dag\phi_c\phi_b^\dag\phi_d)
\,,
}
it becomes custodial-invariant.
This term is contained in the second expression of \eqref{su2lr:inv} since
\eq{
\tfrac{1}{2}\Tr[(\Phi_a^\dag\Phi_b)(\Phi_c^\dag\Phi_d)]=
\re(\phi_a^\dag\phi_b)\re(\phi_c^\dag\phi_d)
-\bz_{ab}\ponto\bz_{cd}\,.
}

Considering the properties $\bz_{ba}=-\bz_{ab}$ and $\bz_{aa}=\bs{0}$,
we note that if some of the indices of \eqref{z.z} are equal, excluding $a=b$ or
$c=d$, the term 
$\im(\phi_a^\dag\phi_b)\im(\phi_c^\dag\phi_d)$ is effectively canceled out.
For example, for $b=c$,
\eq{
\label{z.z:b=c}
\bz_{ab}\ponto\bz_{bd}=
\re(\phi_a^\dag\phi_b)\re(\phi_b^\dag\phi_d)
-\phi_b^\dag\phi_b\re(\phi_a^\dag\phi_d)
\,.
}
Hence terms $\bz_{ab}\ponto\bz_{cd}$ with two equal indices do not generate
new genuine fourth-order custodial invariants besides the product of two quadratic
invariants.

The property \eqref{z.z:b=c} means that custodial-invariant potentials with $N=2$
and $N=3$ Higgs doublets do not contain terms of the form
$\im(\phi_a^\dag\phi_b)\im(\phi_c^\dag\phi_d)$.
The custodial-invariant 2HDM in its canonical form is, in fact, equivalent to a
CP-invariant (real) potential without the quartic term
$\im^2(\phi_1^\dag\phi_2)$\,\cite{deshpande.ma,maniatis:so4,haber:III}.

Therefore, we can state that
\emph{\!
a NHDM potential, with $N\le 3$, invariant by the custodial group $\so4C$ in its
canonical implementation \eqref{Phi:LR} is equal to a potential invariant by CCP
\,\eqref{def:CCP} with the additional elimination of terms of the form
$\im(\phi_a^\dag\phi_b)\im(\phi_c^\dag\phi_d)$.\!
}
From a CP-invariant potential in its real basis, the latter elimination corresponds
to the elimination of one term for $N=2$ and 6 terms for $N=3$.
In this case, it is straightforward to construct a general $\so4C$-invariant
potential: only consider the terms in Eqs.\,\eqref{present:re} and
\eqref{present:re.re}.
Such potential will have $9$ and $27$ real parameters for the 2HDM and 3HDM,
respectively. 
[A total of $2+2(r+q)+(r+q)(r+q+1)/2$ real parameters, where $r=N-1$ and
$q=N(N-1)/2$.]

One important remark concerning the description above is that for the full quantum
theory considering radiative corrections, since gauge interactions do not respect
the full custodial group $\so4C$, the effective potential will almost inevitably
contain terms of the form $\im(\phi_a^\dag\phi_b)\im(\phi_c^\dag\phi_d)$ generated by
finite radiative corrections.
But since CP symmetry is a good symmetry of the potential and gauge interactions,
the terms \eqref{absent:im} and \eqref{absent:re.im} can only be generated from
other sectors that violate CP, such as Yukawa interactions.

For $N\ge 4$, the terms $\im(\phi_a^\dag\phi_b)\im(\phi_c^\dag\phi_d)$ might be
present in the potential contained in new genuine fourth order custodial invariants
such as
\eq{
\label{z.z:1234}
\bz_{12}\ponto\bz_{34}=
\im(\phi_1^\dag\phi_2)\im(\phi_3^\dag\phi_4)
+\re(-\phi_1^\dag\phi_4\phi_2^\dag\phi_3 +\phi_1^\dag\phi_3\phi_2^\dag\phi_4)
\,.
}
We note, however, that the custodial invariants $\bz_{ab}\ponto\bz_{cd}$, with $a\neq
b\neq c\neq d$, still contain quadratic combinations of second-order custodial
invariants $\re(\phi_a^\dag\phi_b)$, and we conclude they are reducible in some way.
We are obviously interested in identifying the irreducible piece.

The irreducible fourth-order invariants can be constructed most easily by using the
real components \eqref{Phi->eta}. We begin by rewriting the second-order invariants
\eq{
\label{inv:2:eta}
\etaa{a}\ponto\etaa{b}=\tfrac{1}{2}\Tr[\Phi_a^\dag\Phi_b]
=\re(\phi_a^\dag\phi_b)
\,.
}
The only $\so4C$ invariant that can be constructed out of four 4-vectors
$\etaa{a},\etaa{b},\etaa{c},\etaa{d}$ which is also independent of \eqref{inv:2:eta}
is the pseudoscalar
\eq{
\label{I:4:eta}
\Iq_{abcd}\equiv
\epsilon_{\mu\nu\alpha\beta}\etaa{a}_\mu\etaa{b}_\nu\etaa{c}_\alpha\etaa{d}_\beta
\,,
}
where the $\epsilon_{\mu\nu\alpha\beta}$ is the totally antisymmetric tensor in
four dimensions obeying $\epsilon_{0123}=1$.
We could have also written $I^{(4)}_{abcd}=\det\etaa{abcd}$ for
$
\etaa{abcd}\equiv
\big(\etaa{a}|\etaa{b}|\etaa{c}|\etaa{d}\big)
$
as a $4\times 4$ matrix.
The $SO(4)$ invariant \eqref{I:4:eta} is the genuine custodial invariant that only
appears for 4 or more vectors (doublets). If any two of the indices $a,b,c,d$ are
equal the invariant is null.

We can rewrite $\Iq$ in terms of Higgs-doublets or $\bz_{ab}$\,\eqref{zi:def} as
\eqarr{
\label{I:4:im}
I^{(4)}_{abcd}&=&
\im(\phi_a^\dag\phi_b)\im(\phi_c^\dag\phi_d)
+\im(\phi_a^\dag\phi_d)\im(\phi_b^\dag\phi_c)
\cr&&
~+\ \im(\phi_a^\dag\phi_c)\im(\phi_d^\dag\phi_b)
\,,\\
&=&
\bz_{ab}\ponto\bz_{cd}
+\bz_{ac}\ponto\bz_{db}
+\bz_{ad}\ponto\bz_{bc}
\,.
}
We note that the imaginary parts of the gauge invariants $\im(\phi_a^\dag\phi_b)$
only appear through the combination \eqref{I:4:im} since
\eqarr{
\bz_{ab}\ponto\bz_{cd}&=&
(\etaa{a}\ponto\etaa{c})(\etaa{b}\ponto\etaa{d})
-(\etaa{a}\ponto\etaa{d})(\etaa{b}\ponto\etaa{c})
\cr&&
+\ I^{(4)}_{abcd}\,.
}

The invariant $\Iq_{abcd}$ changes sign by exchange of any pair of indices
among $(abcd)$ and hence by any odd permutation.
An even permutation of $(abcd)$, on the other hand, leaves $\Iq_{abcd}$ invariant.
Thus each term $\Iq_{abcd}$ is invariant by an $A_4\subset SO(4)_H$ discrete
symmetry in each horizontal subspace spanned by $\{\phi_a,\phi_b,\phi_c,\phi_d\}$.

Another useful formula is
\eq{
\det(\im K^{(1234)})=
(\Iq_{1234})^2\,,
}
where $K^{(1234)}$ is the submatrix of $K$, given by
$K_{ab}=\phi_b^\dag\phi_a$\,\cite{maniatis:1,nhdm:orbit}, constructed from the
intersection of the rows 1,2,3,4 with columns 1,2,3,4.

Notice that the doublets live in the space $\CC^2$ and then at most
two doublets can be linearly independent. Instead, when we are transported to
$\RR^4$ through Eq.\,\eqref{Phi->eta}, four 4-vectors can be linearly independent, as
they should be.
For example, if $\phi_3=\alpha\phi_1+\beta\phi_2$ for $\phi_1,\phi_2$ linearly
independent, then still $\{\etaa{3},\etaa{1},\etaa{2}\}$ is linearly independent
if $\im\alpha\neq 0$ or $\im\beta\neq 0$.

For completeness, we can rewrite the $SU(2)_R$ vectors \eqref{zi:def} as vectors of
$\so3C$:
\eq{
\bz_{ab}=\etaa{a}_0\betaa{b}-\etaa{b}_0\betaa{a}+\betaa{a}{\times}\betaa{b}\,.
}

The most general custodial invariant NHDM potential for $N\ge 4$ can be written as
the most general linear combination of terms of
Eqs.\,\eqref{present:re}, \eqref{present:re.re} and \eqref{I:4:im}.
For the invariants $\Iq_{abcd}$, there is only one invariant for each
set $(abcd)$ with all distinct indices.
For example, for $N=4$, we know there is only one invariant. 
For $N\ge 5$, there will be $\binom{N}{4}$ invariants.

To perform the minimization of the potential with $N\ge 5$, however, we need to know
additionally how many invariants $\Iq$ are functionally independent among the
$\binom{N}{4}$ invariants.

\subsection{Isotropy groups and symmetry breaking}
\label{subsec:vevs}

Let us analyze here what are the possible symmetry groups that are left invariant
when the $N$-Higgs-doublets acquire nonzero vacuum expectation values (VEVs),
required from successful EWSB, in a potential invariant by the custodial group
$\so4C$.
The VEVs are realized as the absolute minimum of the Higgs potential and their
isotropy groups define the spontaneous symmetry-breaking pattern.
The specific patterns depend on the parameters of the potential but one can study
the maximal realizable symmetry-breaking patterns from the possible
isotropy groups of general VEVs with respect to both $\so4C$ (global)
and $\gsm$ (gauge).

Let us begin with the custodial group and consider $N$ Higgs-doublets $\phi_a$,
together with their corresponding natural forms $\Phi_a$ and $\etaa{a}$, upon which
$\sulr{2}$ and $\so4C$ act naturally as in Eqs.\,\eqref{Phi:LR} and \eqref{eta:SO4}
respectively. We want to study how the possible isotropy groups $H_C$ change as $N$
grows.
We will assume that all $\aver{\phi_a}\neq \bs{0}$. Otherwise, it will coincide with
the VEVs for smaller $N$. In any case, successful EWSB also requires
$\sum_a\aver{\phi_a^\dag\phi_a}=v_L^2/2$, where $v_L\sim 246\rm GeV$.

For $N=1$ Higgs doublet $\phi_1$ (SM), $\so4C$ is an automatic global
symmetry of the Higgs potential and the custodial symmetry $\so3C$ always remains
in the potential after EWSB breaking because we can rotate away any corresponding VEV
$\aver{\Phi_1}$ by a global symmetry $\su2L$ alone and obtain
\eq{
\label{vev:Phi1}
\aver{\Phi_1} = v_1\id\,,
}
where $\sqrt{2}v_1=246\rm GeV$ within the SM. 
The isotropy group is $H_C=\su2{L+R}\sim\so3C$.
The VEV above automatically conserves CP.
In terms of the real field $\etaa{1}$ it is also obvious that any VEV
$\aver{\etaa{1}}=(v_1,0,0,0)$ would leave an $\so3C$ group invariant.

For $N=2$ Higgs doublets $\phi_1,\phi_2$, the global symmetry $\so4C$ is not
automatic and has to be imposed in the potential. But once it is imposed,
CP is an automatic symmetry of the potential. 
If we consider that EWSB proceeds through the VEVs
$\aver{\Phi_1}\neq 0,\aver{\Phi_2}\neq 0$, then we can rotate $\aver{\Phi_1}$ to the
form \eqref{vev:Phi1} by a global $\su2L$ transformation.
We can still apply an $\su2{L+R}$ transformation [$U_R=U_L$ in Eq.\,\eqref{Phi:LR}]
which leaves $\aver{\Phi_1}$ invariant but allows us to write
\eq{
\label{vev:Phi2}
\aver{\Phi_2}=
v_2e^{-i\varphi_2\sigma_3}\,.
}
We have parametrized $\aver{\etaa{2}}=(v_2\cos\varphi_2,0,0,-v_2\sin\varphi_2)$ for
convenience.
If $\varphi_2\neq 0,\pi$, the isotropy group left invariant  by $\aver{\Phi_1}$ and
$\aver{\Phi_2}$ in Eqs.\,\eqref{vev:Phi1} and \eqref{vev:Phi2} is
$H_C=U(1)_{T_{3R}+T_{3R}}\sim\so2C$.

For $N\ge 3$ Higgs doublets, the isotropy group can be reduced to the trivial
$H_C=\{\id\}$ or some discrete group.
If $\aver{\Phi_1},\aver{\Phi_2}\neq 0$, we can rotate
$\aver{\Phi_1},\aver{\Phi_2}$ to the forms \eqref{vev:Phi1} and \eqref{vev:Phi2}
through global transformations.
For $\aver{\Phi_3}\neq 0$, we can use the $U(1)_{T_{3R}+T_{3R}}$ freedom to write
\eq{
\label{vev:Phi3}
\aver{\Phi_3}=
\aver{\etaa{3}_0}\id+
\aver{\etaa{3}_2}i\sigma_2+
\aver{\etaa{3}_3}i\sigma_3\,.
}
If $\aver{\etaa{3}_2}\neq 0$, there is no continuous symmetry left and the isotropy
group is $H_C=\{\id\}$ or some discrete group.
For the nonzero VEVs $\aver{\Phi_a}$, $a\ge 4$, all the possible components are
meaningful because there is no continuous symmetry.

In terms of the gauge group $\gsm$, the possible isotropy groups $H$ for
$N=1,2,3,\ldots$, are known.
For $N=1$, $H=\uem$ and the conventional form for $\aver{\phi_1}$ is
\eq{
\label{phi:1}
\aver{\phi_1}= 
\begin{pmatrix}
0 \cr v_1
\end{pmatrix}
\,,\quad
v_1>0\,.
}
It is the equivalent of Eq.\,\eqref{vev:Phi1} in terms of doublets.

For $N=2$, we can preserve the form of \eqref{phi:1}.
The isotropy group $H$ is trivial if $a_2$ is nonzero for the general form
\eq{
\label{vev:phi2:cb}
\aver{\phi_2'}=
v_2
\begin{pmatrix}
a_2 \cr b_2
\end{pmatrix}
\,,\quad
v_2>0,\quad a_2\ge 0,
}
where $b_2$ is a nonzero complex number.
For nonzero $a_2$, this VEV corresponds to charge breaking vacuum (CB-V).
This VEV contrasts to Eq.\,\eqref{vev:Phi2} whose doublet version is
\eq{
\label{vev:phi2:cpv}
\aver{\phi_2}= 
v_2
\begin{pmatrix}
0 \cr  e^{i\varphi_2}
\end{pmatrix}
\,,\quad
v_2>0.
}
It corresponds to a vacuum that spontaneously breaks CP (CPB-V) in a real potential.
In a general 2HDM potential, the VEVs \eqref{vev:phi2:cb} and \eqref{vev:phi2:cpv}
correspond to physically very different situations.
In a custodial invariant potential, however, they are in the same $\so4C$ stratum
and have the same $\so2C$ isotropy group with respect to $\so4C$.

In fact, in a custodial-invariant potential, 
\textit{a CPB vacuum and a CB vacuum are always degenerate at tree level}.
In other words, if a VEV of the form \eqref{vev:phi2:cpv} is a minimum of the
potential, then there are infinite continuously related VEVs of the form
\eqref{vev:phi2:cb} corresponding to the same minimum value of the potential.
The reason is that
$V(U_L\aver{\Phi_1}U_R^\dag,U_L\aver{\Phi_2}U_R^\dag)
=V(\aver{\Phi_1},\aver{\Phi_2})$
and we can relate the matrices $\aver{\Phi_a}$ in
Eqs.\,\eqref{vev:Phi1} and \eqref{vev:Phi2} to the following matrices through an
$\su2{L+R}$ transformation:
\eqarr{
U_L\aver{\Phi_1}U_R^\dag&=&
\aver{\Phi_1}
\,,\\
\label{Phi':2}
U_L\aver{\Phi_2}U_R^\dag&=&
v_2(\cos\varphi_2\id -\sin\varphi_2i\bs{\sigma}\ponto\bn)
\,,
}
where $\bn=(\sin\theta\cos\alpha,\sin\theta\sin\alpha,\cos\theta)$
and 
\eq{
\label{ULR:32}
U_L=U_R=e^{-i\sigma_3\alpha/2}e^{-i\sigma_2\theta/2}\,.
}
The VEV \eqref{Phi':2} corresponds to the form \eqref{vev:phi2:cb} if we
associate $b_2=\cos{\varphi_2}+i\sin\varphi_2\cos\theta$ and
$a_2=\sin\varphi_2\sin\theta$ for $\alpha=3\pi/2$ in Eq.\,\eqref{ULR:32}.

Another way of looking into the degeneracy between a CPB vacuum and a CB vacuum is to
analyze the basis invariant that signals CB\,\cite{nhdm:V}:
$
\aver{r_0}^2-|\aver{\br}|^2\ge 0\,.
$
Only the equality signals a neutral vacuum.
The variables $r_\mu$, $\mu=0,1,\ldots,{\small N^2-1}$ are linear
combinations of $\phi_a^\dag\phi_b$\,\eqref{phi:ab} that will be properly 
defined in Eq.\,\eqref{r.mu}.
In the simplest case of 2HDM, the $r_\mu$ variables are
$2r_0=|\phi_1|^2+|\phi_2|^2$, $2r_3=|\phi_1|^2-|\phi_2|^2$, 
$r_1=\re(\phi^\dag_1\phi_2)$ and $r_2=\im(\phi^\dag_1\phi_2)$, which allow us to
write
\eq{
\label{rmu.rmu}
\aver{r_0}^2-|\aver{\br}|^2
=|\aver{\phi_1}|^2|\aver{\phi_2}|^2
-[\re\aver{\phi_1^\dag\phi_2}]^2
-[\im\aver{\phi_1^\dag\phi_2}]^2\,.
}
Notice that the first two terms are custodial invariants while the third
is not\,\eqref{phi->Phi:im}.
The latter, $\im\aver{\phi_1^\dag\phi_2}=\aver{z_{12}^3}\neq 0$, signals SCPV in the
real basis.
Therefore, if a potential minimum obeys $\aver{r_0}^2-|\aver{\br}|^2= 0$ but
$\im\aver{\phi_1^\dag\phi_2}\neq 0$, one can use $\su2R$ to rotate
$\aver{\bz_{12}}$ in \eqref{zi:def} to decrease the 3-component and then increase 
$\aver{r_0}^2-|\aver{\br}|^2$ in a $\so4C$-invariant manner.

If we recall the identity \eqref{z.z} for $(cd)=(ab)=(12)$,
\eq{
|\bz_{12}|^2=|\phi_1|^2|\phi_2|^2-\re^2(\phi_1^\dag\phi_2)\,,
}
one can see that 
\eq{
\label{z12=0}
|\aver{\bz_{12}}|^2=0
}
signals a neutral and CP-invariant vacuum in a custodial-invariant manner.
In this case, the first two terms and the third term in Eq.\,\eqref{rmu.rmu} are
independently null.

The generalization of Eq.\,\eqref{z12=0} to NHDM is given by
\eq{
\label{zab=0}
\sum_{a<b}|\aver{\bz_{ab}}|^2=0\,.
}
This condition ensures the symmetry-breaking pattern $\so4C\to \so3C$.

If the symmetry-breaking pattern $\so4C\to \so3C$ is realized in 
a NHDM potential, the residual symmetry $\so3C$ guarantees that all the VEVs of
the doublets are aligned as 
\eq{
\label{vev:align}
\aver{\Phi_a}=v_a\id\,,\quad a=1,2,\ldots,N\,,
}
restricted by
\eq{
2\sum_a v^2_a=v_L^2\,,
}
with $v_L\sim246\rm GeV$ from successful EWSB.
The residual symmetry also guarantees that spontaneous CP violation is not possible.

It is important to emphasize that the degeneracy between CB and CPB vacua happens at
tree level.
Radiative corrections coming from the gauge sector and the Yukawa sector may break
the degeneracy in the effective potential due to interactions that are not $\so4C$
symmetric. 
Such tree-level degeneracy is a consequence of the symmetry-breaking pattern
$\so4C\to \so2C$ which generates 5 Goldstone bosons.
If the low-lying vacuum, after radiative corrections, is neutral and then CPB, 
three of them will be absorbed as the longitudinal components of the massive gauge
fields. The remaining two scalars will combine to form a charged
pseudo-Goldstone boson\,\cite{pseudo-G} that acquires a small mass due to radiative
corrections\,\cite{deshpande.ma}.
The same is true for any NHDM potential with only two nonzero VEVs that can be
reduced to \eqref{vev:Phi1} and \eqref{vev:Phi2}.
We are not interested in such a scenario due to the absence of the custodial symmetry
that protects the $\rho$ parameter and the lightness of such charged
pseudo-Goldstone boson.

The other scenario, i.e., the 2HDM scenario with a low-lying CB vacuum (with
radiative corrections) or a NHDM,
with $N\ge 3$, with a third doublet with nonzero VEV obeying
Eq.\,\eqref{vev:Phi3} (tree-level),
has no $\uem$ symmetry and contains one more would-be Goldstone boson that gives
the photon a mass through the Higgs mechanism. This possibility is thus
phenomenologically excluded.

The actual parameter space that avoids the possibility of
realizing $\varphi\neq 0,\pi$ in Eq.\,\eqref{vev:Phi2} or $\aver{\etaa{3}_1}\neq 0$
in Eq.\,\eqref{vev:Phi3} for the global minimum can only be mapped by constructing
and analyzing the potential. For a 2HDM custodial-invariant potential with
additional $Z_2$ symmetry, $\phi_2\to-\phi_2$, both $\so4C\to \so3C$ and 
$\so4C\to \so2C$ symmetry-breaking patterns are realizable and the explicit
parameter ranges for them can be found in Refs.\,\cite{sikivie:NPB80,deshpande.ma}.

\subsection{Basis transformation}
\label{sec:basischange}

Since the $N$ Higgs-doublets have the same gauge quantum numbers, any unitary change
of basis will maintain the form of the kinetic terms and gauge interactions for
the doublets but may modify the form of the potential as well as the Yukawa
interactions without changing the physical content of the theory. It amounts to a
reparametrization of the potential and Yukawa coefficients.

The most general set of unitary transformations that commutes with the gauge group
$\gsm$ is given by the basis transformations
\eq{
\label{phi:basis}
\phi_a\to U_{ab}^H\phi_b,\quad U^H\in SU(N)_H\,.
}
We are already factoring out the global $U(1)$ factor because it corresponds to
$U(1)_Y$ transformations.

We can equally write the basis transformation law \eqref{phi:basis} for 
$\Phi_a$\,\eqref{Phia} as\,\cite{endnote:joao}
\eqarr{
\label{UH:Phi}
\Phi_a&\to& (U_{ab}^{H^*}\tphi_b|U_{ac}^{H}\phi_c)\cr
&&=\re(U_{ab}^{H})\Phi_b
+\im(U_{ab}^{H})\oPhi_b
}
where
\eq{
\label{def:oPhi}
\oPhi_a\equiv i(-\tphi_a|\phi_a)=\Phi_a(-i)\sigma_3\,.
}
Notice $\widetilde{\oPhi}_a=\oPhi_a$ under \eqref{def:tilde:22}.
If $U^H$ is a complex transformation, it is clear that linear combinations of
$\{\phi_a\}$ in the horizontal space of doublets will correspond to linear
combinations of $\{\Phi_a\}$ and $\{\oPhi_a\}$, i.e., only the set $\{\Phi_a\}$ is
not sufficient.
From the definition \eqref{def:oPhi} we see that a complex basis transformation
implicitly induces a transformation in $\su2R$ and it shows that $\su{N}H$ is not
independent of $\su2R$.
The transformation law for $\etaa{a}$ can be easily extracted from
Eq.\,\eqref{UH:Phi}.

As extensively discussed\,\cite{nhdm:cp,botella.silva,silva:symmetries}, basis
transformations might conceal the presence of symmetries by changing how a symmetry
acts.
That is the case of CP symmetry for example\,\cite{nhdm:cp,botella.silva,GH}.
That is also the case of the canonical implementation \eqref{Phi:LR} of the custodial
group $\so4C$.
Hence, it is important to study how the representation \eqref{Phi:LR}
of the custodial group $\so4C$ is viewed in other bases connected by
\eqref{phi:basis}.

To that end, it is useful if we use a representation space where all the groups
in question, i.e., $\so4C$ and $SU(N)_H$, act linearly through one matricial
representation. It is better to treat $\so4C$ as in \eqref{chi:LR}.

If we define a complex column vector of size $4N$ (it lives in a 4$N$-dimensional
\textit{real} vector space) by
\eq{
\label{def:chi:N}
\chi\equiv\begin{pmatrix}
\phi \cr -\tphi
\end{pmatrix}
\,,
}
where
\eq{
\label{def:phi:N}
\phi\equiv\begin{pmatrix}
\phi_1\cr \phi_2\cr \vdots \cr \phi_N
\end{pmatrix}
\,,
}
the canonical representation \eqref{chi:LR} of $\sulr{2}$ translates into
\eq{
\label{LR:chi:N}
\chi \to
(U_R\pd\id_N\pd U_L)\,\chi
\,.
}
Notice $\tphi$ in Eq.\,\eqref{def:chi:N} denotes the application of the
tilde operation on each doublet of $\phi$ in Eq.\,\eqref{def:phi:N}.

If we define new Higgs-doublets $\phi'$ related by a change of
basis~\eqref{phi:basis},
\eq{
\label{phi:U1}
\phi=U_1\phi'\,,\quad
\tphi=U_1^*\tphi'\,,
}
we can see the canonical representation \eqref{LR:chi:N} changes into
\eq{
\begin{pmatrix}
\phi' \cr -\tphi'
\end{pmatrix}
\to 
\begin{pmatrix}
U_1^\dag & \cr & U_1^\tp
\end{pmatrix}
(U_R\pd\id_N\pd U_L)
\begin{pmatrix}
U_1 & \cr & U_1^*
\end{pmatrix}
\begin{pmatrix}
\phi' \cr -\tphi'
\end{pmatrix}
\,.
}

We can see the representations of $SU(2)_L$ is not modified and we will omit such
factor within this analysis. 
The change in the representation of $SU(2)_R$ can be completely specified if we know
the new representation of its generators $\{T'_{kR}\}$:
\eq{
T'_{kR}=
\begin{pmatrix}
U_1^\dag & \cr & U_1^\tp
\end{pmatrix}
T_{kR}
\begin{pmatrix}
U_1 & \cr & U_1^*
\end{pmatrix}
\,,\quad k=1,2,3.
}
Obviously $T_{kR}=\tfrac{1}{2}\sigma_k\otimes\id_N$ for the canonical
implementation and we obtain explicitly
\eqarr{
\label{TkR:U1:explicit}
T'_{1R}&=&
\tfrac{1}{2}
\begin{pmatrix}
0 & U^\dag \cr U & 0
\end{pmatrix}
\cr
T'_{2R}&=&
\tfrac{1}{2}
\begin{pmatrix}
0 & -iU^\dag \cr iU & 0
\end{pmatrix}
\\
\nonumber
T'_{3R}&=&\tfrac{1}{2}\sigma_3\pd\id_N = T_{3R}
\,,
}
where
\eq{
\label{U1:U}
U=U_1^\tp U_1\,.
}

As expected, the representation of $T_{3R}\sim \tfrac{1}{2}Y$ is not modified since
the basis transformations \eqref{phi:basis} preserve the gauge structure.
Only the representation of the generators $T_{1R},T_{2R}$ of the coset space
$\su2R/U(1)_Y$ is modified.

We can also note that the representation for $\su2R$ will be modified only when $U_1$
in Eq.\,\eqref{phi:U1} is complex. In fact, if we restrict the basis transformation
group $\su{N}H$ to $\so{N}H$, $U_1$ is orthogonal, $U=\id_N$ in Eq.\,\eqref{U1:U}
and the (representation of) generators \eqref{TkR:U1:explicit} are the canonical
ones.
The converse is also true: the only subgroup of $\su{N}H$ which preserves the
representation of $\su2R$ in Eq.\,\eqref{TkR:U1:explicit} is the group of matrices 
$U_1$ obeying $U_1^\tp U_1=\id_N$, which is equivalent to $U^*_1=U_1$, i.e., they
form the real subgroup $\so{N}H$.
This fact is also in accordance with Eq.\,\eqref{phi:U1}
which can be rewritten  as $\chi_a=(U_1)_{ab}\chi'_b$ for $U_1\in\so{N}H$.
Hence, there are infinitely many canonical implementations of the custodial group
$\so4C$ related by $\so{N}H$.

We can interpret the previous fact as the realization that 
\textit{
only the real $\so{N}H$ subgroup of the basis transformation group $\su{N}H$ commutes
with the custodial group $\so4C$.
}
This also means that the whole global symmetry of the potential can still be
enlarged by discrete or continuous subgroups of $\so{N}H$ without affecting the
custodial group.

\subsection{Equivalence among different implementations}
\label{sec:equiv}

We will show in this section the following: 
\textit{
Any implementation of the custodial group $\sulr{2}/Z_2$ on a potential containing
$N$ Higgs-doublets corresponds to a canonical implementation \eqref{Phi:LR} in some
basis provided that
}
\begin{enumerate}
\item the Higgs kinetic term is invariant by the custodial group;
\item the association $T_{3R}=\frac{1}{2}Y$ fix $U(1)_Y\subset SU(2)_R$;
\item $\su2R$ acts through a reducible representation of $N$ copies of
the fundamental $\bs{2}$-representation.
\end{enumerate}

We already analyzed the converse in Sec.\,\ref{sec:basischange}, i.e., a canonical
implementation of the custodial symmetry satisfies conditions 1, 2 and 3
in any basis.

To prove the statement, we begin by considering $\chi$ in
Eq.\,\eqref{def:chi:N} as the
appropriate representation space of the $N$ Higgs-doublets $\phi_a$.
It has the same degrees of freedom of considering the real and imaginary parts of
each component of each doublet as independent ($4N$ real fields).
Then we note that the representation of $\su2L$ is fixed by the gauge interactions
and each doublet $\phi_a$ transforms according to the same fundamental representation
$\mathbf{2}$.
If $U_L$ is one element of $\su2L$ in the fundamental representation, it acts
reducibly on $\chi$ as
\eq{
\chi\stackrel{\su2L}{\to} (\id_{2N}\pd U_L)\,\chi\,.
}
The action above defines the explicit representation $D(U_L)=\id_{2N}{\otimes} U_L$.

For the $\su2R$ factor, we want the representation $D(U_R)$ of $U_R\in\su2R$ to
commute with any $U_L\in\su2L$, i.e., $[D(U_L),D(U_R)]=0$ for any $U_L,U_R$.
From the irreducibility of $\bs{2}$ of $\su2L$, the only possibility of the action of
$\su2R$ is
\eq{
\chi\stackrel{\su2R}{\to} (\DNN(U_R){\otimes}\id_2)\,\chi\,.
}
We need to specify the representation $\DNN(U_R)$ in the representation
$D(U_R)=\DNN(U_R)\pd\id_2$.
The following consistency condition should be additionally imposed on such
representation in a later stage:
\eq{
\label{chi->chi*}
(\epsilon\pd\id_N\pd\epsilon)\chi^*=\chi\,.
}
For the $\su2L$ factor the condition above is automatic.
Note that Eq.\,\eqref{chi->chi*} means that complex conjugation acts as a linear
transformation on $\chi$.

We begin now to impose condition (1), which requires the form-invariance of
the Higgs kinetic term
\eq{
\partial_\mu\phi^\dag \partial^\mu\phi
=\tfrac{1}{2}\partial_\mu\chi^\dag \partial^\mu\chi\,,
}
under the global symmetry $\sulr{2}$.
The invariance under $\su2L$ is automatic.
The invariance under $\su2R$ implies the representation $D^{^\mt{\!(2N)}}$ should be
unitary, i.e.,
\eq{
\label{D:unitary}
\Big(\DNN(U_R)\Big)^\dag \DNN(U_R)=\id_{2N}
\,,\quad
\text{for all $U_R\in\su2R$.}
}
Recall that any $\su2{}$ representation is equivalent to a unitary representation 
because it is a compact Lie group.

The representation $\DNN$ of the Lie group $\su2R$ also induces a representation of
the respective Lie algebra $su(2)_R$ through the exponential mapping.
The unitarity of $\DNN$ for $\su2{}$ then induces a Hermitian representation for
$su(2)$.
We can then work on the group and the algebra interchangeably if we are only
interested in the connected part of the group modulo the center.
We know $su(2)$ is a three-dimensional Lie algebra and the Pauli matrices
$\tau_k\equiv\tfrac{1}{2}\sigma_k$, $k=1,2,3$, can be used as a basis of the
$su(2)$ abstract Lie algebra obeying $[\tau_i,\tau_j]=i\epsilon_{ijk}\tau_k$.
We can also regard the explicit form of $\{\tau_k\}$ as the basis for $su(2)$ in the
fundamental representation.
Using the basis $\{\tau_k\}$, a general element in the $su(2)$ real algebra can be
written as $\tau_k a_k$, where $a_k$, $k=1,2,3$, are real coefficients.
The exponential mapping then reads
\eq{
\label{su2:exp}
U_R=\exp(i\tau_k a_k)\,,\quad
\DNN(U_R)=\exp(\DNN(i\tau_k)a_k)\,.
}

We want to embed $U(1)_Y$ into $\su2R$ by requiring condition (2), i.e.,
by associating hypercharge to the unique diagonal generator of $\su2R$:
$T_{3R}=\frac{1}{2}Y$.
The action of hypercharge is fixed by the gauge quantum number assignments
\eq{
\chi\stackrel{U(1)_Y}{\to} e^{i\theta\frac{1}{2}Y}\chi
=(e^{i\theta\tau_3}{\otimes}\id_N{\otimes}\id_2)\,\chi
\,.
}
Thus, within the $N$-Higgs-doublet sector, we obtain
\eq{
T_{3R}=\tau_3{\otimes}\id_N{\otimes}\id_2\,.
}
and
\eq{
\label{D:T3R}
\DNN(\tau_3)=\tau_3{\otimes}\id_N\,.
}
Now we can identify $U(1)_Y\sim U(1)_{T_{3R}}$.
Notice we could have associated $T_{3R}=-Y/2$ and this would only change
Eq.\,\eqref{D:T3R} to $\DNN(\tau_3)=-\tau_3{\otimes}\id_N$, and the rest of the
generators of $\su2R$ accordingly.

We also know all the representations of $\su2R$ are totally reducible and 
our requirement (3) of the presence of $N$ irreducible $\bs{2}$-representations
implies that there is a $2N\vezes 2N$ nonsingular matrix $U'$ for which
\eq{
\label{D:reduced:UR}
U^{\prime -1}\DNN(U_R)U'=U_R\pd\id_N
\,,\quad
\text{for all $U_R\in\su2R$.}
}
In the algebra Eq.\,\eqref{D:reduced:UR} translates into
\eq{
\label{D:reduced:tau}
U^{\prime -1}\DNN(\tau_k)U'=\tau_k{\otimes}\id_N
\,,\quad
\text{for $k=1,2,3$.}
}
Notice the usual block diagonal form 
(direct sum) of $N$ copies of irreducible $\bs{2}$-representations for $\tau_k$ is
$\id_N{\otimes}\tau_k=\diag(\tau_k,\tau_k,\ldots,\tau_k)$. But the latter can be
transformed into the form \eqref{D:reduced:tau} by a similarity transformation.

Now, by using Eqs.\,\eqref{D:T3R} and Eq.\,\eqref{D:reduced:tau}, we obtain the
condition
\eq{
[U',\tau_3\pd\id_N]=0\,.
}
Since $\tau_3\pd\id_N$ is already diagonal, $U'$ has to have a block diagonal
form, each block corresponding to distinct eigenvalues of $\tau_3\pd\id_N$
\eq{
\label{U:U12}
U'=
\begin{pmatrix}
U_1 & 0 \cr
 0 & U_2
\end{pmatrix}\,.
}
The matrices $U_1,U_2$ are nonsingular complex matrices of size $N\vezes N$.
Additionally, the unitarity of the representation $\DNN$ \eqref{D:unitary} or the 
hermiticity of \eqref{D:reduced:tau} implies $U_1,U_2$ have to be unitary matrices.

The consistency condition \eqref{chi->chi*} yields
\eq{
(\epsilon{\otimes}\id_N)
\big(D^{^\mt{\!(2N)}}\!(U_R)\big)^*
(\epsilon{\otimes}\id_N)^\dag
=D^{^\mt{\!(2N)}}\!(U_R)\,.
}
By using \eqref{U:U12} we can rewrite
\eqarr{
\begin{pmatrix}
0 & -\id_N \cr \id_N & 0
\end{pmatrix}
\begin{pmatrix}
U_1^* & 0 \cr 0 & U_2^*
\end{pmatrix}
(U_R^*{\otimes}\id_N)
\begin{pmatrix}
U_1^\tp & 0 \cr 0 & U_2^\tp
\end{pmatrix}
\begin{pmatrix}
0 & \id_N \cr -\id_N & 0
\end{pmatrix}
\cr
=
\begin{pmatrix}
U_1 & 0 \cr 0 & U_2
\end{pmatrix}
(U_R{\otimes}\id_N)
\begin{pmatrix}
U_1^\dag & 0 \cr 0 & U_2^\dag
\end{pmatrix}
\,,~~
}
which reduces to
\eq{
\begin{pmatrix}
U_2^\tp U_1 & 0 \cr 0 & U_1^\tp U_2
\end{pmatrix}
(U_R{\otimes}\id_N)
=
(U_R{\otimes}\id_N)
\begin{pmatrix}
U_2^\tp U_1 & 0 \cr 0 & U_1^\tp U_2
\end{pmatrix}\,.
}
The identity \eqref{tU=U} was assumed.

Since the previous identity has to be true for the algebra as well, we can use
$U_R\to i\tau_1$ from which we conclude that
\eq{
\label{U:12:T}
U_1^\tp U_2=U_2^\tp U_1\,.
}
The same condition arises from $U_R\to i\tau_2$.
The $i$-factor in $i\tau_1,i\tau_2$ should be present to preserve the
condition \eqref{chi->chi*}, which should be valid in the group representation but
the latter is related to the representation of the algebra by the relation
\eqref{su2:exp}.

The explicit representation for $\tau_k$ \eqref{D:reduced:tau} is then
\eqarr{
\label{D:tau:U}
\DNN(\tau_1)&=&
\tfrac{1}{2}
\begin{pmatrix}
0 & U^\dag \cr
U & 0
\end{pmatrix}
\cr
\DNN(\tau_2)&=&
\tfrac{1}{2}
\begin{pmatrix}
0 & -iU^\dag \cr
iU & 0
\end{pmatrix}
\\ \nonumber
\DNN(\tau_3)&=&\tau_3\pd\id_N\cr
}
where 
\eq{
\label{U->U12}
U=U_2 U_1^\dag
\,.
}
Notice the matrices in Eq.\,\eqref{D:tau:U} represent appropriately the algebra
$[\tau_i,\tau_j]=i\epsilon_{ijk}\tau_k$.

We can check $U$ is symmetric by using \eqref{U:12:T}
\eq{
U^\tp=U_1^* U_2^\tp
=U_1^*U_1^\tp U_2U_1^\dag
=U\,.
}
If we compare Eq.\,\eqref{D:tau:U} to Eq.\,\eqref{TkR:U1:explicit}, we conclude
immediately that the matrices in Eq.\,\eqref{D:tau:U} correspond to the generators of
$\su2R$ implemented in some basis $\phi'=U_4^\dag\phi$.
The unitary basis transformation matrix $U_4$ should satisfy 
\eq{
\label{U:U4}
U_4^\tp U_4=U\,,
}
where $U$ is given by \eqref{U->U12}. It was shown in Ref.\,\onlinecite{GH}
(Appendix\,B) that the decomposition \eqref{U:U4} is always possible for 
a unitary symmetric matrix $U$.
Thus, there is always a basis for which any implementation of the custodial
group obeying the requirements 1, 2 and 3 can be cast into the canonical form
\eqref{Phi:LR}.

\subsection{Custodial-invariant potential in the canonical form}

In this section we want to analyze the most general NHDM potential invariant by the
custodial group $\so4C$ in the canonical implementation (CI) and seek
general features.
Since any implementation of $\so4C$ is equivalent to the CI in
some basis, one can always use the latter basis to treat the potential.
There are infinitely many such bases.

We begin by writing the most general renormalizable NHDM as\,\cite{nhdm:cp,nhdm:V} 
\eq{
\label{pot:gen}
V=M_\mu r_\mu +\Lambda_{\mu\nu} r_\mu r_\nu \,,
}
where $r_\mu$ are variables that depend bilinearly on the fields (doublets) and
$\{M_\mu,\Lambda_{\mu\nu}\}$ are coefficients.
The convention of summation of repeated indices is implicit with
Euclidean metric instead of the Minkowski metric used in
Ref.\,\onlinecite{nhdm:cp}. In this way, all the indices referring to the variables
$r_\mu$ will be written as lower indices.

The variables $r_\mu$ are defined as\,\cite{nhdm:cp,nhdm:V}
\eq{
\label{r.mu}
r_\mu\equiv (T_\mu)_{ab}\phi^\dag_a\phi_b\,,
~~\mu=0,1,\ldots,d.
}
The $N\vezes N$ matrices $T_\mu$ are
\eq{
T_0\equiv\sqrt{\frac{N\!-\!1}{2N}}\id_{\mt{N}}\,,\quad
T_i\equiv \frac{1}{2}\lambda_i\,,
}
where $\{\lambda_i\}$ are the $d=N^2-1$ hermitian generators of $SU(N)_H$ in the
fundamental representation, obeying the normalization
$\Tr[\lambda_i\lambda_j]=2\delta_{ij}$.

We can choose $\{T_i\}$ to be the generalization of the Gell-Mann matrices of
$SU(3)$ to $SU(N)$.
We separate them into three classes: diagonal matrices $\{h_i\}$,
off-diagonal symmetric matrices $\{\cS_{ab}\}$, and off-diagonal antisymmetric
matrices $\{\cA_{ab}\}$, with $r,q,q$ elements, respectively [$r=N-1$,
$q=\frac{1}{2}N(N-1)$].
The matrices $\{h_i\}$ correspond to a basis for the Cartan subalgebra while the
$\cS_{ab}$ and $\cA_{ab}$ correspond to the symmetric and antisymmetric
combinations of ladder operators labeled by the pair $(ab)$, $a<b$,
$a,b=1,\ldots,N$.
More explicitly, we can write
\eq{
\cS_{ab}=\tfrac{1}{2}(e_{ab}+e_{ba})
\quad\text{and}\quad
\cA_{ab}=\tfrac{1}{2i}(e_{ab}-e_{ba})
\,,
}
where $e_{ab}$ are the canonical matrices obeying
$(e_{ab})_{ij}=\delta_{ai}\delta_{bj}$.
We can follow, for instance, the ordering 
\eqarr{
\label{order:roots}
(ab)&=&(12),(23),\ldots, (N-1,N),(13),\cr&&
\ldots,(N-2,N),\ldots (1,N)\,.
}
For example, for $SU(3)$ we would have
$\{h_1,h_2\}=\{\lambda_3/2,\lambda_8/2\}$,
$\{\cS_{12},\cS_{23},\cS_{13}\}=\{\lambda_1/2,\lambda_6/2,\lambda_4/2\}$ and
$\{\cA_{12},\cA_{23},\cA_{13}\}=\{\lambda_2/2,\lambda_7/2,\lambda_5/2\}$.
The ordering \eqref{order:roots} follows the ordering of the lower to higher heights
of the roots and the first $r$ of them correspond to the simple roots.

We will denote the variables $r_\mu$ corresponding to $\{h_i,\cS_{ab},\cA_{ab}\}$
respectively by
\eq{
\{\rab{i}, \rab{ab},\rabb{ab}\}\,.
}
Therefore $\{\rab{i}\}$ are combinations of $\phi_a^\dag\phi_a$ while
\eq{
\rab{ab}=\re(\phi_a^\dag\phi_b)\,,\quad
\rabb{ab}=\im(\phi_a^\dag\phi_b)\,,\quad a< b\,.
}

Then the elimination of the terms \eqref{absent:im} and \eqref{absent:re.im}
corresponds in Eq.\,\eqref{pot:gen} to 
\eq{
\label{M:so4}
M_\mu=0 \quad
\text{for $\mu=(\overline{ab})$}
}
and
\eq{
\label{L:so4:re.im}
\Lambda_{\mu\nu}=0 
\quad
\text{for $\mu=(\overline{ab})$ or $\nu=(\overline{ab})$}\,,
}
where \textit{or} means one of them exclusively.
The restriction \eqref{L:so4:re.im} leads to the block diagonal form for the
$d\times d$ submatrix $\tilde{\Lambda}=(\Lambda_{ij})$:
\eq{
\label{Lt:block}
\tilde{\Lambda}=
\begin{pmatrix}
A_{p\times p} & 0_{p\times q} \cr
0_{q\times p} & B_{q\times q} 
\end{pmatrix}
\,,
}
where $p\equiv r+q=N(N+1)/2$ and $0_{p\times q}$ denotes the null matrix of the
shown dimension.

For $N\le 3$, we have to impose additionally 
\eq{
\label{L:so4:im.im}
\Lambda_{\mu\nu}=0 
\quad
\text{for $\mu=(\overline{ab})$ and $\nu=(\overline{ab})$}\,.
}
The restriction \eqref{L:so4:im.im} sets $B_{q\times q}=0_{q\times q}$ in
Eq.\,\eqref{Lt:block}.
The most general custodial invariant 2HDM and 3HDM potentials are then given by
\eqref{pot:gen} with the restrictions \eqref{M:so4}, \eqref{L:so4:re.im} and
\eqref{L:so4:im.im}:
\eq{
\label{pot:gen:ab:3}
V=\sum_{\mu=0,(i),(ab)} M_\mu r_\mu
+\sum_{\mu,\nu=0,(i),(ab)} \Lambda_{\mu\nu}r_\mu r_\nu
\,,
}
where the summation is explicitly shown.

In this case, we can confirm the direct connection between the presence of \emph{only
one charged would-be Goldstone boson} and a \emph{CP-conserving vacuum}.
Such connection follows from the complete absence of the terms
$\im(\phi_a^\dag\phi_b)$ in the potential \eqref{pot:gen:ab:3}, from which we
immediately conclude that independently of $r_\mu$,
\eq{
\label{V:no:im.im}
\frac{\partial V}{\partial \rabb{ab}}=0\,,\quad
a<b=1,\ldots,N\,.
}
Therefore, we can immediately write the extremum equations as\,\cite{nhdm:V}
\eq{
\label{MM}
\frac{\partial V}{\partial \phi_{ai}}=
(\bbM)_{ab}\phi_{bi}=
\sum_{\mu=0,(i),(cd)}
\frac{\partial V}{\partial r_\mu}
(T_{\mu})_{ab}\phi_{bi}
\,.
}
We know $\aver{\bbM}$, the matrix $\bbM$ computed for the minimal values (VEVs), is
the squared-mass matrix for the charged scalars for a neutral vacuum\,\cite{nhdm:V},
including the charged Goldstone bosons.
If there is only one (would-be) charged Goldstone boson, then the VEVs
corresponding to the second components of each doublet,
$\aver{w_a}=\aver{\phi_{a2}}$, form the unique eigenvector of $\aver{\bbM}$
associated with the null eigenvalue.
We can see from Eq.\,\eqref{MM} that $\bbM$ is real and symmetric because the sum
only contains generators $T_\mu$ that are real and symmetric.
Since the eigenvalues of $\aver{\bbM}$ have to be real, the eigenvector $\aver{w}$
is also real and then spontaneous CP violation is excluded\,\cite{endnote:maniatis}.
Obviously, this happens because we are implicitly selecting the symmetry-breaking
pattern $\so4C\to\so3C$. The direct connection above, however, is not apparent for
custodial-invariant NHDM potentials with $N\ge 4$.

The independence of the potential on any $\rabb{ab}$ \eqref{V:no:im.im}
also explains why CB and SCPV vacua are degenerate for the custodial-invariant 2HDM. 
In this case, the potential does not depend on $r_2$, which can be changed
continuously and independently in that direction (within the surface and interior of
the lightcone in Ref.\,\cite{ivanov:mink}) without changing the potential value.
The physical range of the variables $r_\mu$\,\eqref{r.mu} in the 2HDM corresponds
to all the points inside and on the surface of a cone (lightcone) defined by
$r_0^2-\br^2=0$, where $\br^2=r_1^2+r_2^2+r_3^2$.
The VEVs lying on the surface of such a cone correspond to neutral vacua while the
interior points correspond to charge-breaking vacua.
Then, if there is a neutral but CP-violating vacuum $\aver{r_\mu}$ obeying
$\aver{r_0^2-\br^2}=0$, with $\aver{r_2}\neq 0$, the vacuum with the same values of
$r_0,r_1,r_3$ but $\aver{r_2}=0$ lies at the same depth of the potential but it is
charge breaking since $\aver{r_0^2-\br^2}>0$.
The same conclusion can be reached for 3HDM if we can find independent directions to
move in the subspace of $r_2,r_5,r_7$ in the orbit space\,\cite{nhdm:orbit}.

For $N\ge 4$, the most general custodial invariant NHDM potentials are obtained 
by adding to \eqref{pot:gen:ab:3} the term
\eq{
\label{pot:ab:4}
\Delta V_4=
\sum_{a<b<c<d}\lambda_{abcd}\Iq_{abcd}\,.
}
From the definition of $\Iq$ \eqref{I:4:im} we can establish the correspondence
\eq{
\label{L:ab:4}
\lambda_{abcd}=
2\Labbb{ab}{cd}=-2\Labbb{ac}{bd}=2\Labbb{ad}{bc}\,,
}
for $a<b<c<d$.
Equation \eqref{L:ab:4} restricts $\Lambda_{\mu\nu}$ for $\mu,\nu=(\overline{ab})$.
For $N=4$, the expression \eqref{pot:ab:4} contains only one term $\Iq_{1234}$.

Notice that for $N\ge 4$, Eq.\,\eqref{V:no:im.im} is not valid. Because of
Eqs.\,\eqref{pot:ab:4} and \eqref{L:ab:4}, however, the coefficients
$\Lambda_{\mu\nu}$ in the sector of
$\rabb{ab}\rabb{cd}$ is a sparse
matrix. For example, for $N=4$, in the subspace of
$\{\rabb{12},\rabb{13},\rabb{14},\rabb{23},\rabb{24},\rabb{34}\}$, the submatrix
$B_{q\times q}$ in Eq.\,\eqref{Lt:block} can be written as
\eqarr{
\label{L:im.im:4}
B_{6\times 6}&=&
\Big(\Lambda_{(\overline{ab})(\overline{cd})}\Big)=
\lambda_{1234}
\begin{pmatrix}
0_{3\times 3} & B_1 \cr
B_1 & 0_{3\times 3}
\end{pmatrix}\,,\quad
\cr
B_1&=&
\begin{pmatrix}
0 & 0 & 1 \cr
0 & -1 & 0 \cr
1 & 0 & 0 \cr
\end{pmatrix}\,.
}
It is clear that the block matrix in Eq.\,\eqref{L:im.im:4} has eigenvalues
$\pm\lambda_{1234}$, each one with multiplicity 3.
The explicit construction for $N=5$, reveals that the corresponding block matrix has
eigenvalues $0,\lambda,-\lambda$, with multiplicities $4,3,3$, respectively, where
$\lambda$ is a positive number that depends on the 5 parameters $\lambda_{abcd}$,
$a<b<c<d$.

\section{Different implementations in 2HDM}
\label{sec:2hdm}

We analyze in the following the different, but equivalent, implementations of $\so4C$
in the 2HDM potential.
The first two implementations, Secs.\,\ref{subsec:twisted} and \ref{subsec:typeII},
were already considered in the literature but we show in Sec.\,\ref{subsec:other}
that they can be generalized.

\subsection{Twisted implementation}
\label{subsec:twisted}

It can immediately be seen that the ``twisted'' implementation of the custodial
symmetry in Ref.\,\onlinecite{herquet:prl} is just the canonical symmetry
\eqref{Phi:LR} imposed in a basis
\eq{
\label{phi:rephase}
\phi'_a=e^{i\theta_a}\phi_a\,,
\quad a=1,2.
}
It corresponds to a change of basis in Eq.\,\eqref{phi:U1} with
\eq{
U_1^\dag=\diag(e^{i\theta_a})\,,
}
where we can choose $\sum_a\theta_a=0$ from $U(1)_Y$ invariance.

This implementation can be straightforwardly generalized to
$N$-Higgs-doublets with general $N\ge 2$.

For the particular case of 2HDM, two apparently inequivalent implementations of
the custodial group were presented in Ref.\,\onlinecite{herquet:prl}.
The twisted version was shown to allow for the degeneracy between the charged
Higgs $H^\pm$ and the CP-even scalar $H^0$, $M_{H_0}=M_{H^\pm}$.
However, as shown in Sec.\,\ref{sec:equiv}, any implementation of the
custodial group is equivalent to the canonical one in some basis.
In the specific case of the twisted version of Ref.\,\onlinecite{herquet:prl}, 
the required change of basis is the rephasing transformation of \eqref{phi:rephase}.
After the appropriate basis change, the canonical CP transformations automatically
constitute a symmetry of the Higgs potential. Let us call such automatic symmetry
CP1.
Thus the ``twisted'' implementation corresponds to an imposition of an additional
CP2 symmetry, orthogonal to CP1 in the sense of Ref.\,\onlinecite{nhdm:cp}, apart
from the automatic CP1 symmetry. With respect to the CP1 symmetry, the CP2-even
scalar $H^0$ is actually CP1-odd. Obviously, one can choose one of the CP symmetries
to be a true symmetry of the theory by extending it to the Yukawa sector. Only then,
is it possible to define (approximately) the scalar degenerate to $H^\pm$ as
CP-even, the CP symmetry being the CP2 transformation.
A similar comment has been already made in Ref.\,\onlinecite{haber:III}.

\subsection{Type II implementation}
\label{subsec:typeII}

For the specific case of the 2HDM, various apparently different implementations of
the custodial group can be devised.
An implementation called type II\,\cite{pomarol:rho} follows from the action of
$\sulr{2}$ upon
\eq{
\label{M12}
M_{12}\equiv (\tphi_2|\phi_1)\,,
}
instead of the action on $\Phi_1,\Phi_2$.
This implementation was first considered in Ref.\,\onlinecite{pomarol:rho} as
distinct from the canonical one, which was named type I.
However, Refs.\,\cite{maniatis:so4,haber:III} show that they are in fact
equivalent to a canonical implementation in a different basis.

To complement such discussion, let us analyze the equivalence between type II and
canonical implementation from a slightly different perspective.

The matrix $M_{12}$ is a general complex $2\vezes 2$ matrix, different from $\Phi_a$
which obeys \eqref{tPhi=Phi}.
In terms of independent degrees of freedom (real fields), $\Phi_1$ and $\Phi_2$
together has 8 degrees of freedom, the same number as for $M_{12}$.
The action of $\sulr{2}$ on $M_{12}$ in Eq.\,\eqref{M12} is, however, reducible
because of the property \eqref{tU=U}.
The irreducible pieces of $M_{12}$ can be separated, respectively, as the even and
odd parts under the tilde operation \eqref{def:tilde:22}:
\eqarr{
\label{M12:+-}
M_{12}^{(+)}&=& \frac{M_{12}+\tilde{M}_{12}}{\sqrt{2}}\,,\cr		
M_{12}^{(-)}&=& \frac{M_{12}-\tilde{M}_{12}}{\sqrt{2}}\,.
}
We can also think the tilde operation as a linear transformation in the space of
$\chi_a$\,\eqref{def:chi} with two eigenvalues, $\pm 1$, that separates the
irreducible representations of $M_{12}$.

In other words, we can define doublets $\phi_1'$ and $\phi_2'$ associated,
respectively, to
\eqarr{
\Phi_1'&\equiv& \phantom{-i} M_{12}^{(+)}=a_\mu\ee_\mu\,,\cr
\Phi_2'&\equiv& -iM_{12}^{(-)}=b_\mu\ee_\mu \,.
}
or, equivalently,
\eq{
M_{12}=\frac{1}{\sqrt{2}}(a_\mu+ib_\mu)\ee_\mu\,,
}
where $a_\mu$ and $b_\mu$ are 4-vectors similar to $\etaa{a}$.
The explicit relation between the doublets $\phi_a$ in \eqref{M12} and $\phi'_a$ is
given by
\eqarr{
\label{II->I}
\phi_1'&=&\frac{\phi_1+\phi_2}{\sqrt{2}}\,,
\cr
\phi_2'&=&\frac{\phi_1-\phi_2}{\sqrt{2}i}\,.
}

One can see, as expected, that the type II implementation on \eqref{M12} corresponds
to the canonical implementation on $\phi_a'$ through $\Phi_a'$ constructed as
Eq.\,\eqref{Phia}. 
It should be remarked that the basis change in Eq.\,\eqref{II->I} connecting type II
implementation and canonical implementation is not unique because any additional
$\so2H$ transformation on ${\phi'_a}$ still preserves the canonical implementation
of $\so4C$.

\subsection{Other implementations}
\label{subsec:other}

We have shown in Sec.\,\ref{sec:equiv} that all implementations of the
custodial group are equivalent to the canonical implementation in some basis.
For the 2HDM potential we can show another practical way of implementing custodial
symmetry, in clear analogy with type II implementation, without resorting to basis
change.

In this class of implementation we define
\eq{
M_{1'1}\equiv (\tilde{\phi}_1'|\phi_1)\,,
}
where 
\eq{
\phi_1'=c_1\phi_1+c_2\phi_2\,,\quad |c_1|^2+|c_2|^2=1\,.
}
We can easily see $(c_1,c_2)=(1,0)$ [in this case $\Phi_2$ will be also necessary]
and $(c_1,c_2)=(0,1)$ correspond to the
canonical and type II implementations respectively.

The action of $\sulr{2}$ is given by
\eq{
M_{1'1}\to U_LM_{1'1}U_R^\dag\,.
}

By analyzing the infinitesimal transformation of $\su2R$ on $\phi_1$ and $\phi_2$,
we can conclude that such implementation is equivalent to the canonical
implementation on $(\phi_1',\phi_2')$ given by
\eq{
\begin{pmatrix}
\phi_1' \cr \phi_2'
\end{pmatrix}
=
\begin{pmatrix}
c_1 & c_2\cr 
c_2 & -\frac{c_1^*c_2}{c_2^*}
\end{pmatrix}
\begin{pmatrix}
\phi_1 \cr \phi_2
\end{pmatrix}
\,.
}
One can check the previous transformation matrix is unitary and symmetric.

\section{Consequences and conclusions}
\label{sec:consequences}

We have shown in Sec.\,\ref{sec:equiv} that any implementation of the custodial
$\so4C\sim\sulr{2}$ group in a NHDM potential is equivalent to the canonical
implementation in some basis if $\su2L$ is the gauge factor, $U(1)_Y$ is
embedded in $\su2R$, and we require $N$ copies of the doublet representation of
$\su2R$.
On the other hand, we have seen in Sec.\,\ref{subsec:Cinv} that a potential
invariant by $\so4C$ in the canonical implementation (CI) is automatically
CP-invariant in its canonical form (CCP).
Therefore, \emph{any NHDM potential invariant by a global $\so4C$ symmetry is also
automatically CP-symmetric independently of the implementation}.
Moreover, the basis for which $\so4C$ implementation is canonical is
automatically aligned to the basis where the CP transformation on the doublets is
canonical.
If one requires the custodial symmetry $\so3C$ to remain after EWSB, spontaneous CP
violation is automatically excluded as well.

One can then take advantage of such result to use explicit CP invariance as a
necessary criterion to identify global $\so4C$ invariance in a NHDM potential, even
if its manifestation is not explicit.
Several tools based on basis invariant conditions were already developed for
identifying CP-invariance in a general NHDM
potential\,\cite{nhdm:cp,botella.silva,GH}.
One can also make use of the fact that a CP-invariant NHDM potential can be always
written in the so-called \textit{real basis} (or the canonical CP
basis\,\cite{nhdm:cp}) where all the parameters of the potential written in terms of
the gauge invariants $\phi_a^\dag\phi_b$ are real\,\cite{GH}. 
In other words, there is always a basis where the CP symmetry manifests
canonically\,\eqref{def:CCP}.

To find such basis for a general CP-invariant NHDM potential with $N\ge 3$ is a
difficult technical and unsolved problem\,\cite{nhdm:cp}.
For the 2HDM, the change of basis from a general basis to the real basis can be
explicitly given\,\cite{nhdm:cp}. 
For the 3HDM, necessary and sufficient criteria
for CP invariance can be devised but the systematic method to find
the real basis is lacking\,\cite{nhdm:cp}.
In both cases, once we write the CP-invariant potential in its real
form, then identifying if the potential is $\so4C$-symmetric is a straightforward
task because any real basis is also a basis where the action of the $\so4C$ symmetry
is canonical.
Thus, it suffices to check if the potential is entirely written in terms
of $\re(\phi_a^\dag\phi_b)$ in its linear and quadratic combinations, just as in
Eq.\,\eqref{pot:gen:ab:3}. Such a criterion is invariant by the residual $\so3H$
reparametrization freedom that remains.

For the specific case of the 2HDM, basis-invariant conditions to test $\so4C$ can
be written\,\cite{maniatis:so4}.
If we follow the methods of Ref.\,\onlinecite{nhdm:cp}, we can consider the
parameters
$\{\bM=(M_i),\bL_0=(\Lambda_{0i}),\tilde{\Lambda}=({\Lambda}_{ij})\}$ as two
vectors and one rank-2 tensor of $\so3H$, respectively.  These parameters were
defined in Eq.\,\eqref{pot:gen}.
CP transformations then act through ordinary reflection in three dimensions along a
direction $\bkcp$ called CP-reflection direction, which might not be unique.
CP invariance then is equivalent to the existence of $\bkcp$ such that
\eq{
\bkcp\ponto\bM=0\,,\quad
\bkcp\ponto\bL_0=0\,,\quad
\text{$\bkcp$ is an eigenvector of $\tilde{\Lambda}$}\,.
}
In addition, the potential is $\so4C$-invariant if, and only if, 
\eq{
\text{$\bkcp$ is an eigenvector of $\tilde{\Lambda}$ associated to the
eigenvalue 0}\,.
}
Obviously $\det\tilde{\Lambda}$ should be null\,\cite{maniatis:so4}.
The possible directions for $\bkcp$ were given in Ref.\,\onlinecite{nhdm:cp} but when
$\bM,\bL_0$ are not null and nonparallel vectors, $\bkcp=\bM\vezes\bL_0$ is
certainly one CP-reflection direction.

For 3HDMs, a systematic procedure to find the real basis was not completed but we
can still write the necessary and sufficient conditions to have $\so4C$ symmetry.
If the potential is CP-symmetric, there should be a three-dimensional CP-odd subspace
$\mathfrak{t}'_q$ of the adjoint space, the 8-dimensional euclidean space where
$\br=(r_i)$ lives, such that\,\cite{nhdm:cp}
\eq{
\bM\perp \mathfrak{t}'_q\,,\quad
\bL_0\perp\mathfrak{t}'_q\,,\quad 
\text{$\mathfrak{t}'_q$ is an invariant subspace of $\tilde{\Lambda}$.}
}
In the real basis, $\mathfrak{t}'_q$ is spanned by
$\{\lambda_2,\lambda_5,\lambda_7\}$.
The potential is additionally $\so4C$-invariant if, and only if,
\eq{
\text{$\mathfrak{t}'_q$ is contained in the nullspace of $\tilde{\Lambda}$}\,.
}

For $N\ge 4$, a real basis may not coincide with a basis of CI of $\so4C$.
More specifically, CCP invariance (real basis) requires the parameters
$\{M_\mu,\Lambda_{\mu\nu}\}$ to obey some but not all of the restrictions for
$\so4C$ invariance in the CI. 
The restrictions for CCP invariance coincide with Eqs.\,\eqref{M:so4},
\eqref{L:so4:re.im} and \eqref{Lt:block}, but $\so4C$ invariance
additionally requires that $B_{q\times q}$ in Eq.\,\eqref{Lt:block}
should be compatible with the structure in Eq.\,\eqref{pot:ab:4}.

For example, for $N=4$, a CP-invariant potential can be brought to the real basis
where we can identify the block matrix $B_{q\times q}$ in $(\Lambda_{\mu\nu})$.
If the potential is also $\so4C$-invariant, the structure of $B_{q\times q}$ should
be of the form in Eq.\,\eqref{L:im.im:4} (CI) or $\so4H$ equivalent. The group
$\so4H$ appears as the real subgroup of the reparametrization group $\su4H$ which
preserves the representations of $\so4C$ and in fact it is the only
compatible reparametrization group. Now, an element $g$ of the group $\so4H$ acts on
$B_{6\times 6}$ as 
\[
B_{6\times 6} \to D^6(g)B_{6\times 6}D^{6\tp}(g),\quad g\in \so4H\,,
\]
where $D^6$ is the $6$-dimensional adjoint representation of $g$. 
Certainly, from Eq.\,\eqref{L:im.im:4}, $\so4C$ invariance requires 
$B_{6\times 6}$ to have only two eigenvalues $\pm\lambda$ with multiplicity 3 each.
The problem is that such condition is not sufficient to guarantee that there is a
matrix $D^6(g)$ that transform $B_{6\times 6}$ into the form \eqref{L:im.im:4}
because a $\so6{}$ orthogonal matrix is necessary in general.
Therefore, additional conditions are necessary.

In the same way, we can extract the necessary condition for $N=5$: 
$B_{q\times q}$ in the real basis should have eigenvalues $0,\lambda,-\lambda$ with
multiplicities $4,3,3$ respectively. 
Analogous necessary conditions for general $N\ge 4$ could be found by
studying the terms in \eqref{pot:ab:4}.

As for the general problem of constructing the most general NHDM
potential invariant by the global $\so4C$ symmetry, 
we managed to solve it in its simplest form: we can write the most general
custodial-invariant potential before EWSB. The details of EWSB and the parameter
range to achieve the desired symmetry-breaking pattern $\so4C\to\so3C$ will be left
for further work.
For the $Z_2$-symmetric 2HDM, the parameter range is known since
Refs.\,\cite{sikivie:NPB80,deshpande.ma}.

In any case, we confirmed that the phenomenologically interesting symmetry-breaking
pattern is $\so4C\to\so3C$.
In this case, the custodial symmetry $\so3C\sim \su3{L+R}$ remains as an approximate
symmetry at the electroweak scale and protects the $\rho$ parameter of acquiring
large radiative corrections quadratic in the masses of the scalars.

The other viable $\so4C\to\so2C$ symmetry-breaking pattern does not have the
custodial symmetry and its viability in realistic models can not be treated at tree
level because the CP-violating but neutral vacua are degenerate with the
charge-breaking vacua. Only if radiative corrections are taken into account, can we
check which models have one neutral vacuum lying deeper than the charge-breaking
vacua.
This fact was not sufficiently emphasized in previous
works\,\cite{haber:III,deshpande.ma,maniatis:so4}.

Because the symmetry-breaking $\so4C\to\so2C$ is possible and spontaneously
CP-violating and charge-breaking vacua might be degenerate, we can conclude that 
\textit{the imposition of the custodial group $\so4C$ before EWSB and the
requirement of $\so3C$ custodial symmetry are separate conditions in NHDMs} (with
$N>1$).
The imposition of the global $\so4C$ symmetry certainly implies explicit CP
invariance in the NHDM potential but the possibility of having spontaneous CP
violation is still present with the possibility of the pattern $\so4C\to\so2C$.

An interesting question that arises through the study of the global $\so4C$ in
NHDMs is the role played by global symmetries in the potential.
For the 2HDM, with $Z_2$ symmetry, different global symmetries ranging from the
usual $G_{\sm}$ to $SO(8)$ were studied in Ref.\,\onlinecite{deshpande.ma}.
In general, for a global symmetry larger than $\so4C$, (pseudo) Goldstone bosons
appear\,\cite{pseudo-G}.
For example, the global symmetry group $G_{\sm}\otimes SU(2)_H$
is broken down to $U(1)\otimes U(1)_{em}$ generating three would-be Goldstone bosons
and two true Goldstone bosons associated with the breaking of $SU(2)_H$. These true
Goldstone bosons are neutral.
Reference \cite{deshpande.ma} also explored the possibilities of having $SO(4)$,
$SO(4)\pd SO(4)$ and $SO(8)$ as global symmetries. The case of the $SO(4)$
group is the usual custodial symmetry extended to 2HDM. The case of $SO(4)\pd
SO(4)$ is the unlocked case where the custodial symmetry acts independently for both
doublets.
In these cases the global symmetry is not an independent factor but contains the
gauge symmetries of the SM.
A detailed study of possible approximate symmetries in NHDMs is performed in
Ref.\,\onlinecite{osland:nhdm}.

The maximal continuous global symmetry in 2HDMs is $SO(8)$ contained in $O(8)$.
In this case, there is no additional continuous symmetry definable and the
reparametrization group $\su2H$ is also contained in $SO(8)$.
Electroweak symmetry breaking proceeds through only one breaking pattern of the 
global symmetry: $SO(8)$ into $SO(7)$,
where 7 Goldstone bosons emerge, among them some will be absorbed by the Higgs
mechanism while the remaining will be pseudo-Goldstone bosons.
The gauge interactions respects only $G_{\sm}$.
Then, the breaking pattern of the gauge group $G_{\sm}$ is not defined at the
tree level of the potential because one nontrivial $SO(8)$ orbit contains both
VEVs that are neutral and charge breaking with respect to $G_{\sm}$. 
Radiative corrections of the gauge interactions should be incorporated to decide
which model preserves $U(1)_{em}$ and to calculate the radiative masses of the
pseudo-Goldstone bosons.
The fate of the CP symmetry may depend on the radiative corrections of the gauge
sector as well as of the Yukawa sector that violates CP.

Therefore the consideration of $\so4C$ as a global symmetry in NHDMs has two
advantages: (i) it can preserve the approximate custodial symmetry $\so3C$ and
(ii) it avoids the presence of pseudo-Goldstone bosons and degenerate symmetry
breaking patterns at tree level.
If we insist on considering only the global $\so4C$ symmetry in NHDMs, the immediate
question towards generalization is if we can use different representations
$n_a$ other than the doublet representation for $\su2R$.
The canonical implementation uses $n_a=2$ for all $\phi_a$.
We immediately conclude that the association $Y/2=T_{3R}$ is not possible for any
other representation $n_a>2$, since we have only two hypercharges, one for $\phi_a$
and the opposite for $\phi_a^*$, common to all the doublets. 
For example, to have $n_a=3$ we need a state with zero hypercharge and to have
$n_a=4$, we need two different hypercharges in modulus.
Related to that, one can obviously embed the electroweak gauge group into a larger
group that is broken at high energies to a model containing two or
more Higgs doublets with a global accidental symmetry inherited
from the high-scale symmetry\,\cite{ma.ng}.

Within the setting of Sec.\,\ref{sec:equiv}, different but equivalent
implementations of $\so4C$ were considered for the 2HDM in the literature.
We reviewed them in Sec.\,\ref{sec:2hdm} and showed a generalization in
Sec.\,\ref{subsec:other}. We explicitly showed the basis change necessary to
obtain the canonical implementation.

Another interesting general feature of the NHDMs implementing the global $\so4C$
symmetry is the role played by the residual $\so{N}H$ reparametrization freedom.
For $N=2$, it is just the freedom to rotate the plane perpendicular to $\bkcp$. For
$N=3$, it is the $\so3H$ freedom to reparametrize $\mathfrak{t}'_q$ and its
orthogonal complement.
Such reparametrization freedom can still be used to study general $\so4C$ invariant
potentials. Another possible role for $\so{N}H$ is the possibility to promote
the whole group or some subgroup $G'_H$ to symmetries of the potential independently
of the $\so4C$ symmetry. The global symmetry in this case would be $\so4C\pd G'_H$.
Obviously, the symmetry $G'_H$ could be broken by EWSB.

It is important to emphasize that the whole horizontal group $\su{N}H$ in NHDM no
longer commutes with the $\so4C$ symmetry and some invariants that are
$\su{N}H$-invariant may not be $\so4C$-invariant. That is the case of the invariant
$\aver{r_0^2-\br^2}$ in 2HDM\,\cite{ivanov:mink,nhdm:V} that characterizes
neutral vacuum if its null. Since it is not an $\so4C$ invariant, its consideration
alone can not distinguish among some symmetry-breaking patterns of the gauge group
for the pattern $\so4C\to\so2C$. A $\so4C$-invariant measure for $\so4C\to\so3C$ was
shown in Eq.\,\eqref{zab=0}.

Another context where basis transformations are important is the connection with the
Yukawa sector.
Although all the implementations of $\so4C$ symmetry are equivalent within the Higgs
potential, some properties of the whole theory (Lagrangian), including symmetries,
might depend on how the symmetry is implemented in one sector relative to the other.
Since the global $\so4C$ is explicitly broken by the gauge interactions and Yukawa
interactions, the breaking effects depend on the implementation of the symmetry in
the potential relative to the other breaking
sectors\,\cite{maniatis:so4,silva:yukawa1}.

Understanding how the breaking of some global symmetries occurs is also important in
other contexts such as flavor physics. For example, Ref.\,\onlinecite{buras:fcnc}
compares the possibility of FCNC in multi-Higgs models under the perspective of the
two most popular suppression mechanisms, namely, Natural Flavor Conservation (NFC)
and Minimal Flavor Violation (MFV)\,\cite{MFV:02}. It is also possible to constrain
2HDM models from the requirement of mass hierarchy and supression of
FCNC\,\cite{blechman}.

In summary, we have shown that all implementations of the $\so4C$ global symmetry in
NHDMs are equivalent within some general conditions such as the presence of $N$ equal
representations of $\su2R$. 
Such results have enabled us to study a general $\so4C$-invariant potential in its
canonical implementation where the CP symmetry is manifest.
We have shown that for $N=2,3$ the difference between $\so4C$-invariant and
CP-invariant potentials is minimal and the knowledge of the CP symmetry immediately
gives information about $\so4C$-invariance or violation in the potential.
We have explicitly reviewed some implementations for the 2HDM and showed a
general implementation.

\acknowledgments
This work was partially supported by Brazilian
{\em Conselho Nacional de Desenvolvimento Científico e Tecnológico} (CNPq)
and 
{\em Fundação de Amparo à Pesquisa do Estado de São Paulo} (Fapesp).

\appendix
\section{Custodial symmetry: SM and notation}
\label{subsec:notation}

The Higgs potential in SM possesses a global accidental $\so4C\sim\sulr{2}/Z_2$
symmetry due to the presence of one isolated doublet representation. Such symmetry
group contains the electroweak gauge group $G_{\sm}\equiv \gsm$ as a subgroup, which
is the true symmetry of the Higgs Lagrangian, including the gauge interactions.

The electroweak symmetry $G_\sm$ is then broken down to $\uem$ by the 
Higgs VEV $\aver{\phi^0}$. At the same time, the EWSB mechanism breaks $\so4C$
down to the subgroup $\so3C\sim \su2{L+R}$ containing $\uem$, which is the 
only nontrivial symmetry-breaking pattern possible.
In the absence of $U(1)_Y$ interactions ($g_Y\to 0$ limit), the three
would-be Goldstone bosons absorbed by the Higgs mechanism form a triplet under
$\so3C$.
Such symmetry protects the $\rho$ parameter from large radiative corrections
quadratic in the Higgs mass and it is called custodial
symmetry\,\cite{sikivie:NPB80,georgi:EFT}.

Instead of distinguishing the various symmetries above by names, we denote them
by their group structure and use the isomorphic groups interchangeably, often
considering their local structure only.
The groups in question are
\eqarr{
\gsm &\subset&
\sulr{2}/Z_2\sim\so4C
\cr
\label{groups}
\downarrow\mss{\rm EWSB}~ && \hs{3.5em}\downarrow\mss{\rm EWSB}
\\
\uem\quad &\subset&
\hs{2.8em}\su2{L+R}\sim\so3C
\,.
\nonumber
}
The subgroup/group relation in the first line of Eq.\,\eqref{groups} also considers
$U(1)_Y \subset \su2R$.

\section{Generators of $\so4C$}
\label{ap:chi->eta}

The transformation matrix $\mathcal{B}$ in \eqref{chi->eta} is given by
\eq{
\mathcal{B}=\begin{pmatrix}
0 & i & 1 & 0  \cr
1 & 0 & 0 & -i  \cr 
-1& 0 & 0 & -i  \cr 
0 &-i & 1 & 0 
\end{pmatrix}\,.
}
This relation can be read from Eq.\,\eqref{Phi->eta}.

If we define the generators of $\su2{L+R}$ and $\su2{L-R}$ acting on $\chi_a$ as 
\eqarr{
L_j&\equiv& T_{Lj}+T_{Rj}=
\id_2{\otimes}\frac{\sigma_j}{2}+\frac{\sigma_j}{2}{\otimes}\id_2
\cr
K_j&\equiv& T_{Lj}-T_{Rj}=
\id_2{\otimes}\frac{\sigma_j}{2}-\frac{\sigma_j}{2}{\otimes}\id_2
}
the generators acting on $\etaa{a}$ are
\eqarr{
L'_j&=& \mathcal{B}^{-1}L_j\mathcal{B}
\cr
K'_j&=& \mathcal{B}^{-1}K_j\mathcal{B}\,.
}
Or explicitly,
\dcol{
\eqarr{
L'_1&=&\begin{pmatrix}
0 & 0 & 0 & 0 \cr
0 & 0 & 0 & 0 \cr 
0 & 0 & 0 & -i \cr 
0 & 0 & i & 0 
\end{pmatrix}
\cr
L'_2&=&\begin{pmatrix}
0 & 0 & 0 & 0 \cr
0 & 0 & 0 & i \cr 
0 & 0 & 0 & 0 \cr 
0 & -i & 0 & 0
\end{pmatrix}
\nonumber
\\
\nonumber
L'_3&=&\begin{pmatrix}
0 & 0 & 0 & 0 \cr
0 & 0 & -i & 0 \cr 
0 & i & 0 & 0 \cr 
0 & 0 & 0 & 0 
\end{pmatrix}
}
}{
\eqarr{
K'_1&=&\begin{pmatrix}
0 & -i & 0 & 0 \cr
i & 0 & 0 & 0 \cr 
0 & 0 & 0 & 0 \cr 
0 & 0 & 0 & 0
\end{pmatrix}
\cr
K'_2&=&\begin{pmatrix}
0 & 0 & -i & 0 \cr
0 & 0 & 0 & 0 \cr 
i & 0 & 0 & 0 \cr 
0 & 0 & 0 & 0
\end{pmatrix}
\\
\nonumber
K'_3&=&\begin{pmatrix}
0 & 0 & 0 & -i \cr
0 & 0 & 0 & 0 \cr 
0 & 0 & 0 & 0 \cr 
i & 0 & 0 & 0
\end{pmatrix}\quad
}
}

We can immediately notice that $\{L_j\}$ are the generators of $\so3C$ and $\{K_j\}$
are the generators of the cosets $\so4C/\so3C$.



\begin{thebibliography}{99}

\bibitem{veltman.97}
  S.~Weinberg,
  ``A Model of Leptons,''
  Phys.\ Rev.\ Lett.\  {\bf 19 } (1967)  1264-1266; 
  M.~J.~G.~Veltman,
  ``Reflections on the Higgs system,''
  CERN Yellow Report {\bf 97-05} (1997).

\bibitem{sikivie:NPB80}
  P.~Sikivie, L.~Susskind, M.~B.~Voloshin and V.~I.~Zakharov,
  ``Isospin Breaking In Technicolor Models,''
  Nucl.\ Phys.\  B {\bf 173} (1980) 189.

\bibitem{georgi:EFT}
  H.~Georgi,
  ``Effective field theory,''
  Ann.\ Rev.\ Nucl.\ Part.\ Sci.\  {\bf 43} (1993) 209.
\bibitem{willenbrock}
  S.~Willenbrock,
  ``Symmetries of the standard model,''
  arXiv:hep-ph/0410370 (lectures presented at TASI 2004).
\bibitem{isidori:2009}
  G.~Isidori,
  ``Effective theories of electroweak symmetry breaking,''
  PoS C {\bf D09} (2009) 073
  [arXiv:0911.3219 [hep-ph]].

\bibitem{lee:scpv}
T. D. Lee,
``A Theory Of Spontaneous T Violation,''
Phys. Rev. D 8, 1226 (1973);
``CP Nonconservation And Spontaneous Symmetry Breaking,''
Phys. Rep. 9, 143 (1974);
S. Weinberg,
``Gauge Theory Of CP Violation,''
Phys. Rev. Lett. 37, 657 (1976).

\bibitem{IDM:dark}
  L.~Lopez Honorez, E.~Nezri, J.~F.~Oliver and M.~H.~G.~Tytgat,
  ``The inert doublet model: An archetype for dark matter,''
  JCAP {\bf 0702} (2007) 028
  [arXiv:hep-ph/0612275].

\bibitem{barbieri.hall}
  R.~Barbieri, L.~J.~Hall and V.~S.~Rychkov,
  ``Improved naturalness with a heavy Higgs: An alternative road to LHC
  physics,''
  Phys.\ Rev.\  D {\bf 74} (2006) 015007
  [arXiv:hep-ph/0603188].

\bibitem{osland}
  B.~Grzadkowski, O.~M.~Ogreid, P.~Osland, A.~Pukhov and M.~Purmohammadi,
  ``Exploring the CP-Violating Inert-Doublet Model,''
  arXiv:1012.4680 [hep-ph].
  B.~Grzadkowski, O.~M.~Ogreid and P.~Osland,
  ``Natural Multi-Higgs Model with Dark Matter and CP Violation,''
  Phys.\ Rev.\  D {\bf 80} (2009) 055013
  [arXiv:0904.2173 [hep-ph]];
  B.~Grzadkowski and P.~Osland,
  ``Tempered Two-Higgs-Doublet Model,''
  Phys.\ Rev.\  D {\bf 82} (2010) 125026
  [arXiv:0910.4068 [hep-ph]].

\bibitem{pilaftsis}
  A.~Pilaftsis, C.~E.~M.~Wagner,
  ``Higgs bosons in the minimal supersymmetric standard model with explicit CP
violation,''
  Nucl.\ Phys.\  {\bf B553 } (1999)  3-42.
  [hep-ph/9902371].

\bibitem{carena}
  M.~Carena and H.~E.~Haber,
 ``Higgs boson theory and phenomenology. ((V)),''
  Prog.\ Part.\ Nucl.\ Phys.\  {\bf 50} (2003) 63
  [arXiv:hep-ph/0208209].

\bibitem{haber.silva:susy}
  P.~M.~Ferreira, H.~E.~Haber and J.~P.~Silva,
  ``Basis invariant conditions for supersymmetry in the two-Higgs-doublet
  model,''
  Phys.\ Rev.\  D {\bf 82} (2010) 016001
  [arXiv:1004.3292 [hep-ph]].

\bibitem{botella.silva}
F.~J.~Botella and J.~P.~Silva,
 ``Jarlskog-like invariants for theories with scalars and fermions,''
  Phys.\ Rev.\ D {\bf 51}, 3870 (1995)
  [arXiv:hep-ph/9411288];
  L.~Lavoura and J.~P.~Silva,
 ``Fundamental CP violating quantities in a SU(2) x U(1) model with many Higgs
 doublets,''
  Phys.\ Rev.\ D {\bf 50}, 4619 (1994)
  [arXiv:hep-ph/9404276].

\bibitem{GH}
  J.~F.~Gunion and H.~E.~Haber,
 ``Conditions for CP-violation in the general two-Higgs-doublet model,''
  Phys.\ Rev.\ D {\bf 72}, 095002 (2005)
  [arXiv:hep-ph/0506227].

\bibitem{haber:I.II}
  S.~Davidson and H.~E.~Haber,
  ``Basis-independent methods for the two-Higgs-doublet model,''
  Phys.\ Rev.\  D {\bf 72} (2005) 035004
  [Erratum-ibid.\  D {\bf 72} (2005) 099902]
  [arXiv:hep-ph/0504050];
  H.~E.~Haber and D.~O'Neil,
  ``Basis-independent methods for the two-Higgs-doublet model. II: The
  significance of tan(beta),''
  Phys.\ Rev.\  D {\bf 74} (2006) 015018
  [arXiv:hep-ph/0602242]
[Erratum-ibid. D {\bf 74} (2006) 059905].

\bibitem{maniatis:1}
  M.~Maniatis, A.~von Manteuffel, O.~Nachtmann and F.~Nagel,
  ``Stability and symmetry breaking in the general two-Higgs-doublet model,''
  Eur.\ Phys.\ J.\  C {\bf 48} (2006) 805
  [arXiv:hep-ph/0605184].

\bibitem{nhdm:cp}
  C.~C.~Nishi,
  ``CP violation conditions in N-Higgs-doublet potentials,''
  Phys.\ Rev.\  D {\bf 74} (2006) 036003
  [Erratum-ibid.\  D {\bf 76} (2007) 119901 (E)]
  [arXiv:hep-ph/0605153].

\bibitem{nhdm:V}
  C.~C.~Nishi,
  ``The structure of potentials with $N$ Higgs doublets,''
  Phys.\ Rev.\  D {\bf 76} (2007) 055013
  [arXiv:0706.2685 [hep-ph]];
  ``Physical parameters and basis transformations in the Two-Higgs-Doublet
  model,''
  Phys.\ Rev.\  D {\bf 77} (2008) 055009
  [arXiv:0712.4260 [hep-ph]].

\bibitem{nhdm:orbit}
  I.~P.~Ivanov and C.~C.~Nishi,
  ``Properties of the general NHDM. I. The orbit space,''
  Phys.\ Rev.\  D {\bf 82} (2010) 015014
  [arXiv:1004.1799 [hep-th]].

\bibitem{nhdm:orbit2}
  I.~P.~Ivanov,
  ``Properties of the general NHDM. II. Higgs potential and its symmetries,''
  JHEP {\bf 1007} (2010) 020
  [arXiv:1004.1802 [hep-th]].

\bibitem{ivanov:mink}
  I.~P.~Ivanov,
  ``Minkowski space structure of the Higgs potential in 2HDM,''
  Phys.\ Rev.\  D {\bf 75} (2007) 035001
  [Erratum-ibid.\  D {\bf 76} (2007) 039902 (E)]
  [arXiv:hep-ph/0609018];
  ``Minkowski space structure of the Higgs potential in 2HDM: II. Minima,
  symmetries, and topology,''
  Phys.\ Rev.\  D {\bf 77} (2008) 015017
  [arXiv:0710.3490 [hep-ph]].


\bibitem{pomarol:rho}
  A.~Pomarol and R.~Vega,
  ``Constraints on CP violation in the Higgs sector from the rho parameter,''
  Nucl.\ Phys.\  B {\bf 413} (1994) 3
  [arXiv:hep-ph/9305272].

\bibitem{herquet:prl}
  J.~M.~Gerard and M.~Herquet,
  ``A twisted custodial symmetry in the two-Higgs-doublet model,''
  Phys.\ Rev.\ Lett.\  {\bf 98} (2007) 251802
  [arXiv:hep-ph/0703051].

\bibitem{haber:III}
  H.~E.~Haber and D.~O'Neil,
  ``Basis-independent methods for the two-Higgs-doublet model III: The
  CP-conserving limit, custodial symmetry, and the oblique parameters S, T,
  U,''
  Phys.\ Rev.\  D {\bf 83} (2011) 055017
  [arXiv:1011.6188 [hep-ph]].

\bibitem{maniatis:so4}
  B.~Grzadkowski, M.~Maniatis and J.~Wudka,
  ``Note on Custodial Symmetry in the Two-Higgs-Doublet Model,''
  arXiv:1011.5228 [hep-ph].

\bibitem{scalar.rho:cancel}
  D.~Toussaint,
  ``Renormalization Effects From Superheavy Higgs Particles,''
  Phys.\ Rev.\  D {\bf 18} (1978) 1626;
  P.~H.~Chankowski, T.~Farris, B.~Grzadkowski, J.~F.~Gunion, J.~Kalinowski and
M.~Krawczyk,
  ``Do precision electroweak constraints guarantee e+ e- collider discovery  of
  at least one Higgs boson of a two Higgs doublet model?,''
  Phys.\ Lett.\  B {\bf 496} (2000) 195
  [arXiv:hep-ph/0009271].

\bibitem{deshpande.ma}
  N.~G.~Deshpande and E.~Ma,
  ``Pattern Of Symmetry Breaking With Two Higgs Doublets,''
  Phys.\ Rev.\  D {\bf 18} (1978) 2574.

\bibitem{pseudo-G}
  S.~Weinberg,
  ``Approximate symmetries and pseudoGoldstone bosons,''
  Phys.\ Rev.\ Lett.\  {\bf 29} (1972) 1698.

\bibitem{silva:symmetries}
  P.~M.~Ferreira and J.~P.~Silva,
  ``Discrete and continuous symmetries in multi-Higgs-doublet models,''
  Phys.\ Rev.\  D {\bf 78} (2008) 116007
  [arXiv:0809.2788 [hep-ph]].

\bibitem{silva:yukawa1}
  P.~M.~Ferreira and J.~P.~Silva,
  ``Generalized CP Invariance and the Yukawa sector of Two-Higgs Models,''
  Eur.\ Phys.\ J.\  C {\bf 69} (2010) 45
  [arXiv:1001.0574 [hep-ph]].

\bibitem{maniatis:yukawa}
  M.~Maniatis, A.~von Manteuffel and O.~Nachtmann,
  ``A new type of CP symmetry, family replication and fermion mass
  hierarchies,''
  Eur.\ Phys.\ J.\  C {\bf 57} (2008) 739
  [arXiv:0711.3760 [hep-ph]];
  M.~Maniatis and O.~Nachtmann,
  ``On the phenomenology of a two-Higgs-doublet model with maximal CP symmetry
  at the LHC,''
  JHEP {\bf 0905} (2009) 028
  [arXiv:0901.4341 [hep-ph]].

\bibitem{endnote:joao}
The author is grateful to João P. Silva for discussions about this issue.

\bibitem{endnote:maniatis}
The author is grateful to Markos Maniatis for discussions about this issue.

\bibitem{osland:nhdm}
  K.~Olaussen, P.~Osland and M.~A.~Solberg,
  ``Symmetry and Mass Degeneration in Multi-Higgs-Doublet Models,''
  arXiv:1007.1424 [hep-ph].

\bibitem{ma.ng}
  E.~Ma and D.~Ng,
  ``Symmetry remnants: Rationale for having two Higgs doublets,''
  Phys.\ Rev.\  D {\bf 49} (1994) 569
  [arXiv:hep-ph/9301245].

\bibitem{buras:fcnc}
  A.~J.~Buras, M.~V.~Carlucci, S.~Gori and G.~Isidori,
  ``Higgs-mediated FCNCs: Natural Flavour Conservation vs. Minimal Flavour
  Violation,''
  JHEP {\bf 1010} (2010) 009
  [arXiv:1005.5310 [hep-ph]];
  F.~J.~Botella, G.~C.~Branco and M.~N.~Rebelo,
  ``Minimal Flavour Violation and Multi-Higgs Models,''
  Phys.\ Lett.\  B {\bf 687} (2010) 194
  [arXiv:0911.1753 [hep-ph]].

\bibitem{MFV:02}
  G.~D'Ambrosio, G.~F.~Giudice, G.~Isidori and A.~Strumia,
  ``Minimal flavour violation: An effective field theory approach,''
  Nucl.\ Phys.\  B {\bf 645} (2002) 155
  [arXiv:hep-ph/0207036].

\bibitem{blechman}
  A.~E.~Blechman, A.~A.~Petrov and G.~Yeghiyan,
  ``The flavor puzzle in multi-Higgs models,''
  JHEP {\bf 1011} (2010) 075
  [arXiv:1009.1612 [hep-ph]].



\end{thebibliography}
\end{document}